\documentclass[sigconf]{acmart}
\usepackage{ragged2e} 
\usepackage{booktabs,makecell, multirow, tabularx, enumitem}

\AtBeginDocument{%
  \providecommand\BibTeX{{%
    \normalfont B\kern-0.5em{\scshape i\kern-0.25em b}\kern-0.8em\TeX}}}

\copyrightyear{2023}
\acmYear{2023}
\setcopyright{acmlicensed}\acmConference[SIGIR '23]{Proceedings of the 46th International ACM SIGIR Conference on Research and Development in Information Retrieval}{July 23--27, 2023}{Taipei, Taiwan}
\acmBooktitle{Proceedings of the 46th International ACM SIGIR Conference on Research and Development in Information Retrieval (SIGIR '23), July 23--27, 2023, Taipei, Taiwan}
\acmPrice{15.00}
\acmDOI{10.1145/3539618.3591646}
\acmISBN{978-1-4503-9408-6/23/07}

\begin{document}


\title{\ourmethod{}: Twitter Bot Detection with Community-Aware\\ Mixtures of Modal-Specific Experts}


\author{Yuhan Liu}
\email{lyh6560@stu.xjtu.edu.cn}
\affiliation{%
  \institution{Xi'an Jiaotong University}
  \streetaddress{28 Xianning W Rd}
  \city{Xi'an}
  \state{Shaanxi}
  \country{China}}

\author{Zhaoxuan Tan}
\email{tanzhaoxuan@stu.xjtu.edu.cn}
\affiliation{%
  \institution{Xi'an Jiaotong University}
  \streetaddress{28 Xianning W Rd}
  \city{Xi'an}
  \state{Shaanxi}
  \country{China}}

\author{Heng Wang}
\email{wh2213210554@stu.xjtu.edu.cn}
\affiliation{%
  \institution{Xi'an Jiaotong University}
  \streetaddress{28 Xianning W Rd}
  \city{Xi'an}
  \state{Shaanxi}
  \country{China}}

\author{Shangbin Feng}
\email{shangbin@cs.washington.edu}
\affiliation{%
  \institution{University of Washington}
  \streetaddress{}
  \city{Seattle}
  \state{WA}
  \country{USA}}

\author{Qinghua Zheng}
\email{qhzheng@mail.xjtu.edu.cn}
\affiliation{%
  \institution{Xi'an Jiaotong University}
  \streetaddress{28 Xianning W Rd}
  \city{Xi'an}
  \state{Shaanxi}
  \country{China}}

\author{Minnan Luo}
\authornote{Corresponding author: Minnan Luo, School of Computer Science and Technology, Xi’an Jiaotong University, Xi’an 710049, China.}
\email{minnluo@xjtu.edu.cn}
\affiliation{%
  \institution{Xi'an Jiaotong University}
  \streetaddress{28 Xianning W Rd}
  \city{Xi'an}
  \state{Shaanxi}
  \country{China}}

\renewcommand{\shortauthors}{Yuhan and Zhaoxuan, et al.}

\begin{abstract}

    Twitter bot detection has become a crucial task in efforts to combat online misinformation, mitigate election interference, and curb malicious propaganda. However, advanced Twitter bots often attempt to mimic the characteristics of genuine users through feature manipulation and disguise themselves to fit in diverse user communities, posing challenges for existing Twitter bot detection models. To this end, we propose \ourmethod{}\footnote{The code is available at \url{https://github.com/lyh6560new/BotMoE}}, a Twitter bot detection framework that jointly utilizes multiple user information modalities (metadata, textual content, network structure) to improve the detection of deceptive bots. Furthermore, \ourmethod{} incorporates a community-aware Mixture-of-Experts (MoE) layer to improve domain generalization and adapt to different Twitter communities. Specifically, \ourmethod{} constructs modal-specific encoders for metadata features, textual content, and graph structure, which jointly model Twitter users from three modal-specific perspectives. We then employ a community-aware MoE layer to automatically assign users to different communities and leverage the corresponding expert networks. Finally, user representations from metadata, text, and graph perspectives are fused with an expert fusion layer, combining all three modalities while measuring the consistency of user information. Extensive experiments demonstrate that \ourmethod{} significantly advances the state-of-the-art on three Twitter bot detection benchmarks. Studies also confirm that \ourmethod{} captures advanced and evasive bots, alleviates the reliance on training data, and better generalizes to new and previously unseen user communities.
    
\end{abstract}



\newcommand{\ourmethod}[1]{}
\renewcommand{\ourmethod}[1]{\texttt{BotMoE}}

\maketitle
\begin{figure}[t]
  \centering
	\includegraphics[width=1.0\linewidth]{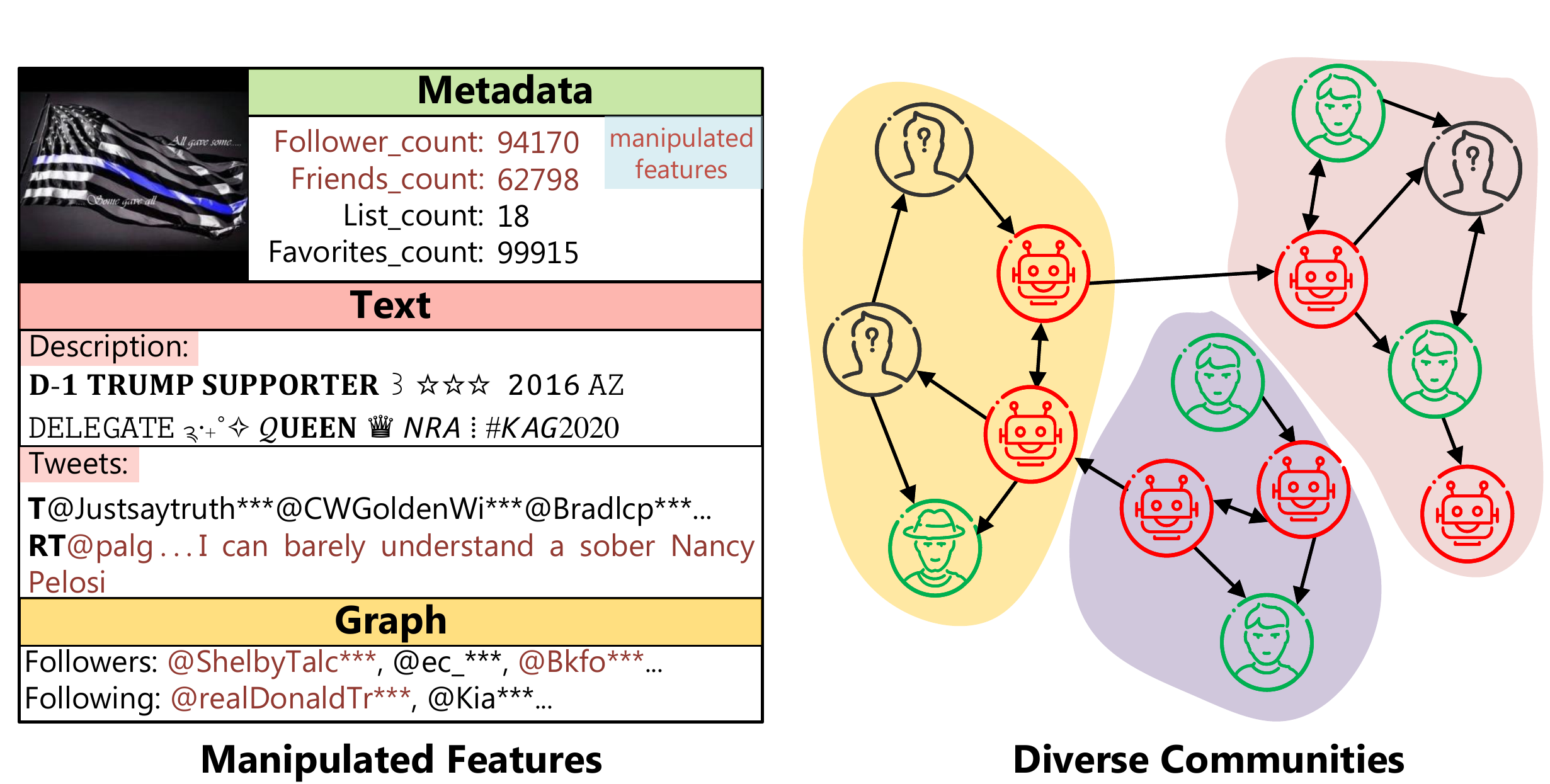}
	\caption{
        \emph{Manipulated features} and \emph{diverse communities} are two challenges facing existing Twitter bot detection systems. All three user information modalities of advanced bots are manipulated to some extent (left), while Twitter bots are widely present in diverse Twitter communities (right).
	}
	\label{fig:teaser}
\end{figure}
\section{Introduction}

Twitter bots, also known as Twitter accounts controlled by automatic programs, have become a widely known phenomenon for social media users \cite{cresci2020decade,ferrara2022twitter}. Twitter bots are posing threats and challenges to online communities by spreading misinformation \cite{cresci2023demystifying,yang2021covid,zannettou2019disinformation,starbird2019disinformation}, deepening societal divides \cite{uyheng2020bots, flores2022datavoidant}, promoting conspiracy theories \cite{ferrara2020covid,greve2022online}, and influencing major elections \citep{ferrara2017disinformation,rossi2020detecting}. As a result, extensive research efforts have been made to counter the malicious impact of Twitter bots through automatic Twitter bot detection systems \cite{davoudi2020towards,chavoshi2016debot}, context-specific case studies \cite{pastor2020spotting,brachten2017strategies}, and more.


Existing automatic Twitter bot detection models could be characterized by their information sources (feature-based, text-based, and graph-based \cite{feng2022twibot}), reflecting the arms race between bot operators and bot detection research \cite{cresci2020decade}: Early Twitter bots are mostly simple spam bots posting repetitive and malicious content \cite{cresci2017paradigm}, thus they could be easily identified by feature-based models through feature engineering and traditional classifiers \cite{yang2020scalable,kudugunta2018deep}. However, in part due to the awareness of these feature-based detectors, bot operators began to tamper with user metadata \cite{cresci2020decade,cresci2017paradigm}, leaving feature-based approaches less effective. As a result, text-based methods were proposed to leverage natural language processing(NLP) techniques \cite{wei2019twitter,dukic2020you,heidari2020using} to analyze the textual content in tweets and user descriptions, aiming to identify Twitter bots through their malicious content. However, text-based methods have become less effective as bots began to post tweets stolen from genuine users \cite{cresci2020decade}, diluting the malicious content in the timeline of Twitter bots. These advanced bots are better captured by graph-based models \cite{li2022botfinder, hu2020heterogeneous,feng2022heterogeneity,lei2022bic}, which leverage graph neural networks to analyze Twitter bot clusters and reveal their collective actions.

Though Twitter bot detection research has come a long way to counter increasingly advanced and evasive Twitter bots, the complexity of bot disguises and the diversified nature of Twitter communities as shown in Figure \ref{fig:teaser} remain underexplored in automatic bot detection systems. As a result, existing Twitter bot detection approaches fall short of addressing these two key challenges:
\begin{itemize}[leftmargin=*]
    \item \textbf{Manipulated Features.} In addition to promoting malicious content, advanced Twitter bots manipulate their information in one or more modalities to evade detection. \citep{kumarage2023stylometric}Deliberately engineering user metadata \cite{cresci2020decade,doi:10.1177/1461444820942744}, copying user descriptions from genuine users \cite{cresci2020decade}, and exploiting follow relationship with the help of botnets \cite{ferrara2022twitter,yang2023fedack} are among the numerous attempts to manipulate user features in metadata, textual content, graph structure, and more. These multi-modal feature manipulation strategies in turn incapacitate the ability of existing bot detectors that merely take one or few information sources into account.
    \item \textbf{Diverse Communities.} The distribution of Twitter bots is heterogeneous across the whole Twitter network \cite{peng2022domain,tan2023botpercent}, while bot behavior and characteristics greatly vary from one user community to another \cite{sayyadiharikandeh2020detection}. However, existing bot detectors often assume the uniformity of Twitter bots, presenting seemingly general-purpose approaches that neglect the heterogeneity of user communities and the difference in bot behavior across different bot clusters. This limitation harms the domain generalization ability of existing bot detectors, creates model fairness issues for employing them for content moderation, and more.
\end{itemize}

In light of these challenges, we propose \ourmethod{}, a novel Twitter bot detection framework that employs a community-aware mixture-of-experts architecture with different experts modeling multi-modal user information sources. Specifically, we first employ modal-specific encoders to jointly leverage the metadata, textual content, and network structure of Twitter users, alleviating the feature manipulation challenge caused by merely analyzing one user information modality. Each modality is then processed by a mixture of modal-specific experts with each expert responsible for the analysis of one specific Twitter community, resulting in community-aware representation learning. These learned representations are further aggregated by assigning Twitter users to one community and activating the corresponding expert. Finally, user representations from each mixture of experts are further synthesized with a novel \emph{Expert Fusion layer}, aiming to fuse input modalities and assess the consistency across the three modalities to detect feature manipulations. The multi-modal user representations and feature consistency matrices are then linearized and used for the binary classification of Twitter bots and genuine users.

 
Extensive experiments demonstrate that \ourmethod{} achieves state-of-the-art performance on three Twitter bot detection benchmarks, significantly outperforming 10 representative baseline models from different categories. In addition, modality fusion study and case study testifies to \ourmethod{}'s ability to tackle manipulated features through jointly leveraging and enabling the interaction across multiple user information modalities. In the user community study and generalization study, we demonstrate that \ourmethod{} better generalizes to new and emerging Twitter communities through the community-aware mixture of experts.

\section{Related Work}
\subsection{Twitter Bot Detection}
Twitter bots are Twitter accounts controlled by automated programs. Bots with malicious goals often harm the integrity of online discourse and reduce search quality. Existing Twitter bot detection methods can be often categorized into three types: feature-based, text-based, and graph-based approaches \citep{feng2021twibot}.

\paragraph{Feature-based methods} 
These methods conduct feature engineering based on handcrafted user features derived from metadata \cite{yang2020scalable,hays2023simplistic,wu2023botshape} and user textual information \cite{kudugunta2018deep}. These features are then combined with traditional classification algorithms \citep{morstatter2016new} to identify Twitter bots. \citet{kudugunta2018deep} exploits user metadata and proposes SMOTENN to combat imbalanced datasets. \citet{miller2014twitter} leverages the feature of tweet content and performs anomaly detection to identify bots. \citet{hayawi2022deeprobot} use hybrid types of numerical, binary, and textural features as high dimensional tensors for classification. However, as noted by \citet{cresci2020decade}, evolving bots can evade the detection of feature-based approaches by creating deceptive accounts with manipulated metadata and stolen tweets.

\paragraph{Text-based methods} 
The methods use NLP techniques to detect Twitter bots with their tweets and descriptions. LSTMs \citep{kudugunta2018deep,luo2020deepbot}, the attention mechanism \cite{feng2021satar}, word embeddings \cite{wei2019twitter,wu2023bottrinet}, and pretrained language models \citep{dukic2020you,silva2019empirical} are leveraged to process tweets for detection. \cite{wei2019twitter} use word vectors and bidirectional LSTMs to process user textual information for bot detection and \citet{dukic2020you} present a BERT-based bot detection model to analyze user tweets. However, text-based methods are easily deceived when advanced bots post stolen tweets and descriptions from genuine users \cite{cresci2020decade}.

\paragraph{Graph-based methods} 
These methods attempt to identify bots based on the graph structure on the Twittersphere. \citet{dehghan2022detecting} use node centrality, while \citet{GraphHist} makes use of a graph's latent local features. Methods leveraging heterogeneous graph neural networks (GNNs) \citep{ali2019detect,feng2021botrgcn,lei2022bic,feng2022heterogeneity,shi2023over,alothali2023sebd} construct a heterogeneous graph based on following, follower, and other user relationships, and learn user representations through graph neural network and achieve state-of-the-art bot detection performance. However, with bot operators creating botnets to manipulate the user networks, methods leveraging only graph information fall short. \citep{cresci2020decade}.

Among these models, there are preliminary attempts to partially address the \emph{manipulated feature} and \emph{diverse community} challenges in Twitter bot detection. For example, BotRGCN \citep{feng2021botrgcn} considers metadata and text features as the initial features for graph convolution. However, shallow concatenation of information sources falls short of identifying advanced bots through compare and contrast while identifying the inconsistency across different modalities. In addition, BotBuster \citep{ng2022botbuster} proposes to train different experts for different features in various modalities and use the combined results of expert outputs. However, this approach assumes the uniformity of bot behavior across communities and platforms, neglecting the heterogeneous nature of Twitter bot distributions.
To this end, we propose \ourmethod{}, a Twitter bot detection framework with community-aware mixtures of modal-specific experts, aiming to improve robustness towards feature manipulation and generalization towards diversified Twitter communities.
 \begin{figure*}[ht]
	\centering
	\includegraphics[width=0.95\linewidth]{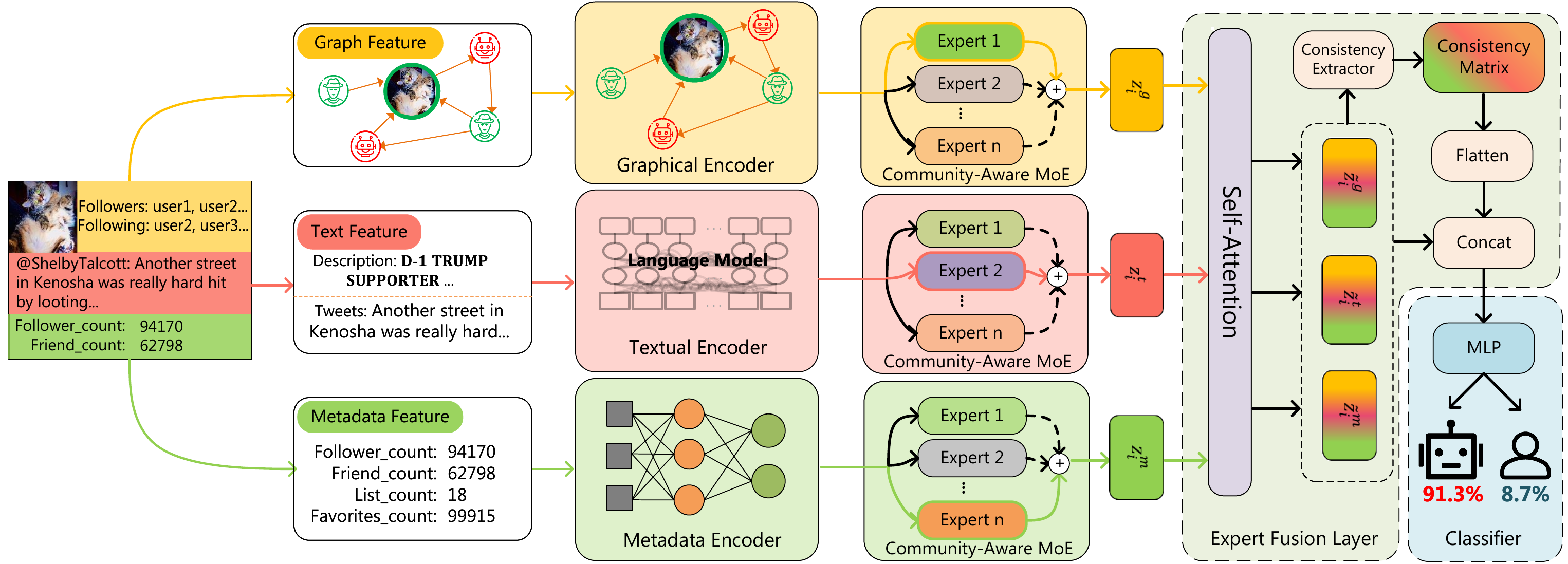}
	\caption{\ourmethod{}: a community-aware Twitter bot detection framework that jointly utilizes users’ metadata, text, and graph features to combat feature manipulation in advanced bots. The metadata, text, and graph information of Twitter users are first processed by encoders, followed by modal-specific community-aware MoE layers. The expert fusion layer is then employed to integrate the three information sources, assess information consistency across modalities, and classify Twitter users.}
	\label{overview}
\end{figure*}
\subsection{Mixture of Experts}
Based on the Divide-and-Conquer principle, the Mixture of Experts (MoE) method is proposed to divide an input sample into sub-tasks and train experts specialized in each sub-task.
 MoE is adopted in NLP to enhance models' capacity \cite{shazeer2017outrageously}, reasoning \cite{madaan2021think}, and generalization \cite{ma2018modeling}. \citet{shazeer2017outrageously} propose a sparsely-gated Mixture-of-Experts layer for conditional computing in large language models. Besides, MoE is also well explored in graph mining: \citet{hu2021graph} use MoE to combat imbalance in graph classification tasks. \citet{fedus2021switch} create simplified routing algorithms for MoE to achieve high training stability and less computational costs. Moreover, MoE can be seen as a natural fit for multi-task learning with different modalities. \citet{ma2018modeling} come up with a multi-gate Mixture-of-Experts with shared experts as sub-models and different gates for each task, aiming to capture inter-task relationships while optimizing task-specific objectives. \citet{bao2021vlmo} use a mixture-of-modality-experts transformer to unify vision and language in training. 

The most relevant work among all MoE-related literature would be BotBuster \citep{ng2022botbuster}, which proposes a multi-platform bot detection framework based on MoE, where each expert is assigned a specific type of user feature to deal with incomplete data. However, BotBuster falls short of leveraging different experts as representatives of different Twitter communities while still assuming the uniformity of Twitter bots across different bot clusters. Therefore, we propose \ourmethod{}, which, in each modality, trains specialized experts for different sub-communities, achieving a community-aware mixture of experts to tackle the \emph{diverse community} challenge.


\section{Methodology}



 Figure \ref{overview} presents an overview of our proposed \ourmethod{} framework with multi-modal user information and community-aware MoE layers. Specifically, we first encode users' metadata, text, and graph information to obtain more comprehensive user representations from three different perspectives. To deal with diverse user communities, we then leverage the Mixture-of-Expert architecture to assign users to experts based on their communities for community-aware bot detection. After that, to enable information interaction, we fuse the user representations from metadata, text, and graph modality with an extractor fusion layer while measuring their inconsistency. Finally, we classify Twitter users into bots or genuine users and learn model parameters.

\subsection{Feature Encoder}
\paragraph{Graph encoder}
To consider the diverse relationships found in social networks, we build a heterogeneous information network (HIN) that captures the follow relationships within the Twittersphere.
 Users' initial features are composed of numerical metadata, categorical metadata, descriptions, and posted tweets to obtain comprehensive user features. We then adopt graph neural networks (GNNs) to encode users' graph modality information while modeling the interaction among users with the message-passing mechanism. The graph-based module in \ourmethod{} can be adapted to various GNN architectures, and their message-passing paradigms can be defined as: 
\begin{align}
\mathbf{a}_{i,r}^{(k+1)} & = \mathrm{AGGREGATE}_{r}^{(k+1)}(\{\mathbf{h}_{j}^{(k)}, \forall j \in \mathcal{N}_{r}(i)\},\mathbf{h}_{\textit{i}}^{(k)}) \\
\mathbf{h}_{i}^{(k+1)} & = \mathrm{UPDATE}^{(k+1)}(\mathbf{h}_{i}^{({k})},\mathbf{a}_{i}^{(k+1)})
\end{align}
where $\mathbf{h}_{i}^{(k)}$ denotes the hidden representation for the user $i$ in layer $k$, $\mathcal{N}_r(i)$ denotes the neighborhood of user $i$, $\mathbf{a}_{\textit{i,r}}^{(k+1)}$ denotes the messages passed from user $i$'s neighborhood under relation $r$, $\mathrm{AGGREGATE}_r$ and $\mathrm{UPDATE}$ denote the aggregation function under relation $r$ and the update function.
We use RGCN \cite{schlichtkrull2018modeling} for the aggregation and update functions due to empirical performance, while our approach could generalize to other GNN architectures such as GCN \cite{kipf2016semi} and RGT \cite{feng2022heterogeneity}.
After $l$ layers of GNN encoders, we obtain the node features $\mathbf{h}_{i}^{l}$ and denote them as the output of the graph encoder $\mathbf{x}_{i}^{g}$ for node $i$.

\paragraph{Textual encoder}

We encode the textual content on Twitter, specifically user tweets and descriptions, with pretrained language models. We leverage pretrained RoBERTa \cite{liu2019roberta} and T5 \cite{raffel2020exploring} as the encoder of the textual module, which can be defined as:
\begin{align}
\mathbf{t}_i=\frac{1}{n}\sum_{a=1}^{n}\frac{1}{l_a}\sum_{b=1}^{l_a}\mathrm{LM}(\mathbf{t}_{i})^{a,b}
\end{align}
where $\mathrm{LM}$ denotes the pretrained language model encoder, $t^{a,b}_{i}$ denotes the $\textit{b}$-th token in $\textit{a}$-th tweet or description for user $i$, $n$ denotes the total number of text content, $l_a$ denotes the length of the input text sequence for $a$-th text content. We then concatenate the features extracted from user descriptions $\mathbf{t}_d$ and tweets $\mathbf{t}_t$ and feed them into a two-layer MLP to get low-dimensional user text representation $\mathbf{x}_{i}^{t}$.

\paragraph{Metadata encoder}
For metadata from user profiles, we consider 5 numerical (followers, followings, statuses, active days, screen name length) and 3 categorical metadata features (protected, verified, default profile image) based on availability and simplicity. After conducting z-score normalization, we apply a two-layer MLP to learn representations for user metadata $\mathbf{x}_{i}^{m}$.
\subsection{Community-Aware MoE Layer}
Accounts from different user communities often have varying characteristics, posing challenges to existing Twitter bot detection systems.
Inspired by the success of MoE architecture in language modeling \cite{peng2020mixture} and machine translation \cite{shazeer2017outrageously}, we take advantage of its ``divide and conquer'' strategy by selectively activating experts on a per-user basis and construct a community-aware MoE layer to process diverse user communities on social networks. Specifically, user representations from metadata, text, and graph modality are first fed into a gate network $G$, which can be illustrated as:
\begin{align}   
G(\mathbf{x})= \mathrm{softmax}(\mathrm{KeepTopK}(\mathbf{W}_g\mathbf{x},k))
\end{align}
where $\mathbf{W}_g \in \mathbb{R}^{d\times n}$ denotes learnable parameters, $n$ denotes the number of experts, $\mathrm{KeepTopK}$ denotes a function to select top k highest gate values given input feature $\mathbf{x}$ \citep{shazeer2017outrageously}. The gating network $G$ produces the output which assigns a user to $k$ different communities and gives the corresponding probability of assignment. The user representations from metadata, text, and graph modality are then fed into their assigned expert networks to conduct community-aware bot detection:
\begin{align}
    \mathbf{z}_{i}^{{\textit{mod}}} = \sum_{j=1}^{n}G(\mathbf{x}_{i}^{\textit{mod}})_j E_{j}(\mathbf{x}_{i}^{\textit{mod}})
    \label{eq:1}
\end{align}
where $G(\mathbf{x}_{i}^{\textit{mod}})_j$ denotes the probability of assigning user $i$ to $j$-th community under $\textit{mod}$ modality and $E_{j}(\mathbf{x}_{i}^{\textit{mod}})$ denotes the output of the $j$-th expert network as a two-layer MLP, $\textit{mod} \in \{m, t, g\}$ denotes the MoE module for metadata, text, and graph modality respectively. The final output $\mathbf{z}_{i}^{m}$ is a weighted sum of the outputs from all selected experts by gate values. Through the MoE architecture, each user is assigned to their corresponding community for processing, enabling the model to make adjustments for the varied user distribution within different communities, thereby addressing the diverse user communities challenge in Twitter bot detection. 


\subsection{Expert Fusion Layer}
After obtaining user representations processed by the experts of the corresponding community, we then combine the representations from the graph, text, and metadata modalities of the same user using a multi-head transformer encoder $\mathrm{TRM}$ to achieve modality interaction,
\begin{align}
\{\Tilde{\mathbf{z}}_{i}^{g},\Tilde{\mathbf{z}}_{i}^{t},\Tilde{\mathbf{z}}_{i}^{m}\}= \mathrm{TRM}(\{\mathbf{z}_{i}^{g},\mathbf{z}_{i}^{t},\mathbf{z}_{i}^{m}\}).
\end{align}
 In addition to the modality interaction, we also evaluate the consistency between graph, text, and metadata representations, aiming to tackle the manipulated features challenge.
 \begin{align}
     \mathbf{z}_{i}^{\textit{con}} = \mathrm{flatten}(\mathrm{sample}(\mathbf{M}) *\omega)
 \end{align}
 Specifically, we perform downsampling on the attention matrix $\mathbf{M}$ with fixed pooling, perform 2D convolution with filter $\omega$, and flatten the output into a vector $ \mathbf{z}^{\textit{con}}_{i}$. In this module, we simultaneously leverage users' metadata, text, and graph information, while measuring the consistency among these modalities at the same time. This multi-modal joint detection paradigm can robustly detect bot accounts with manipulated features. 



\subsection{Training and Inference}
After fusing graph, text, and metadata features and measuring their consistency, we obtain the final user representations for each modality.
We then concatenate the fused representations and the consistency vector $ \mathbf{z}_{i}^{\textit{con}}$, and apply a linear transformation to obtain bot detection result $\mathbf{y}_i$:
\begin{align}
\mathbf{y}_i=\mathbf{W}_{o} \cdot [\Tilde{\mathbf{z}}_{i}^{g},\Tilde{\mathbf{z}}_{i}^{t},\Tilde{\mathbf{z}}_{i}^{m},\mathbf{z}_{i}^{\textit{con}}] + \mathbf{b}_{o},
\end{align}
where $\mathrm{W}_{o}$ and $\mathbf{b}_{o}$ are learnable parameters, $[\cdot, \cdot]$ denotes the concatenation operation.

To train \ourmethod{}, we adopt the cross-entropy loss with a $L_2$ regularization term to combat overfitting. We also employ a balancing loss $\mathrm{BL}(\cdot)$ for each MoE layer to force a fair load and importance among experts. The overall loss function can be presented as:

\begin{align}
\textit{Loss}=-\sum_{i\in Y}t_{i} \log{y_{i}} +\lambda_1 \sum_{\omega \in \theta} \omega^{2}+ \lambda_2 \sum_{i\in Y}\sum^{\{g,t,m\}}_{\textit{mod}}BL(\mathbf{x}_{i}^{\textit{mod}})
\end{align}
where $y_{\textit{i}}$ is predicted output for $i$-th user , $t_{\textit{i}}$ is the corresponding ground truth label, $\theta$ represents all trainable model parameters, and $\lambda_1$ and $\lambda_2$ are hyperparameters maintaining the balance among the three. For balancing loss $\mathrm{BL}$, we refer to \citet{shazeer2017outrageously} to encourage each expert to receive a balanced sample of users:


\begin{align}
BL(x)=w_{\textit{imp}} \cdot \mathrm{CV}(G(x))^{2} + w_{\textit{ld}} \cdot \mathrm{CV}(P(x,i))^{2}
\end{align}
where $\mathrm{CV}$ denotes the coefficient of variation \citep{everitt1998cambridge}, $G(x)$ denotes gate network output, $P(x, i)$ is the smooth function defined in \citet{shazeer2017outrageously}, $w_{\textit{imp}}$ and $w_{\textit{ld}}$ are hyperparameters that balance expert importance and load.

\section{Experiment}

\renewcommand\arraystretch{0.6}
\begin{table}[t]
    \caption{Hyperparameter settings of \ourmethod{}.\label{tab:hyper}}
    \centering   
    \resizebox{1\linewidth}{!}{
    \begin{tabular}{l|ccc}
    \toprule[1.5pt]
    \textbf{Hyperparameter} & \textbf{Cresci-15} & \textbf{TwiBot-20} & \textbf{TwiBot-22}\\
    \midrule[0.75pt]
    optimizer & Adam & Adam & Adam \\
    learning rate &$10^{-4}$ &$10^{-4}$ &$10^{-4}$ \\
    $L2$ regularization $\lambda$ &$10^{-6}$ &$10^{-6}$ &$10^{-6}$\\
    $BL$ coefficient $w_{exp}$ & $10^{-2}$ &$10^{-2}$ &$10^{-2}$\\
    dropout & 0.3 & 0.3 & 0.3\\
    hidden size & 256 & 256 &128 \\
    maximum epochs & 400 & 400 & 400 \\
    transformer attention head $C$ & 2 & 2 & 2 \\
    experts for graph & 2 & 2 & 3\\
    experts for text & 2 & 2 &2\\
    experts for metadata & 3 & 3 & 3\\  
    \bottomrule[1.5pt]   
    \end{tabular}
    }  
\end{table}

\begin{table*}[t]
    \centering
    \caption{Accuracy and binary F1-score of Twitter bot detection systems on the three datasets. We run each method five times and report the average value as well as the standard deviation in parentheses. \textbf{Bold} indicates the best performance, \underline{underline} the second best, and `-' indicates the method is not scalable to TwiBot-22 \cite{feng2022twibot}. * denotes that the results are significantly better than the second-best under the student t-test. \ourmethod{} significantly outperforms all 10 feature-, text-, and graph-based baselines on the three datasets.}
    \begin{tabular}{l c c c c c c }
         \toprule[1.5pt]
         \multirow{2}{*}{\textbf{Model}} & \multicolumn{2}{c}{\textbf{Cresci-15} } & \multicolumn{2}{c}{\textbf{TwiBot-20}} & \multicolumn{2}{c}{\textbf{TwiBot-22}}\\
         \cmidrule(r){2-3} \cmidrule(r){4-5} \cmidrule(r){6-7}
         & Accuracy & F1-score  & Accuracy & F1-score & Accuracy & F1-score   \\ 
         \midrule[0.75pt]
        SGBot \cite{yang2020scalable}    & $77.10~(\small{1.1})$  & $77.91~(\small 0.1)$  & $81.60~(\small 0.5)$   & $84.90~(\small 0.4)$  
                     & $75.10~(\small 0.1)$  & $36.59~(\small 0.2)$     \\
        {Wei
        \textit{et al.} \cite{wei2019twitter}}        & $96.10~(\small 0.1)$  & $82.65~(\small 2.2)$   & $71.30~(\small 0.2)$  & $57.33~(\small 3.2)$  
                     & $70.20~(\small 0.1)$  & $53.61~(\small 1.4)$    \\
        {RoBERTa} \cite{liu2019roberta}       & $96.90~(\small 0.0)$  & $95.86~(\small 0.2)$   & $75.50~(\small 0.0)$  & $73.09~(\small 0.6)$  
                     & $72.10~(\small 0.0)$  & $20.53~(\small 1.7)$\\
        {Varol
        \textit{et al.}} \cite{varol2017online}        & $93.20~(\small 0.0)$  & $94.73~(\small 0.4)$   & $78.70~(\small 0.0)$  & $81.08~(\small 0.5)$  
                     & $73.90~(\small 0.0)$  & $27.54~(\small 0.3)$     \\
        {SATAR} \cite{feng2021satar}  & $93.40~(\small 0.5)$  & $95.05~(\small 0.3)$   & $84.00~(\small 0.8)$  & $86.07~(\small 0.7)$  
                     & -      & -         \\
        {Botometer} \cite{yang2022botometer}    & $57.90$            & $66.90$             & $53.10$            & $53.13 $           
                     & $49.90$  & $42.75$    \\
        {GraphHist} \cite{GraphHist}    & $77.40~(\small 0.2)$  & $84.47~(\small 8.2)$   & $52.10~(\small 0.0)$  & $67.56~(\small 0.3)$  
                     & -      & -    \\
        {BotRGCN} \cite{feng2021botrgcn}       & $96.50~(\small 0.7)$  & $97.30~(\small 0.5)$   & $85.70~(\small 0.7)$  & $87.25~(\small 0.7)$  
                     & $\underline{78.87}~(\small 0.3)$  & $\underline{54.99}~(\small 2.3)$    \\
        {RGT} \cite{feng2022heterogeneity}           & $\underline{97.20}~(\small 0.0)$  & $\underline{97.78}~(\small 0.2)$   & $\underline{86.60}~(\small 0.0)$  & $\underline{88.01}~(\small 0.4)$  
                     & $76.50~(\small 0.0)$  & $42.94~(\small 1.9)$    \\
        {BotBuster} \cite{ng2022botbuster}    & $96.90~(\small 0.3)$  & $97.53~(\small 0.3)$   & $77.24~(\small 0.3)$  & $81.18~(\small 0.4)$  
                     & $74.06~(\small 1.2)$  & $54.18~(\small 7.0)$    \\
        \midrule[0.75pt]
        \textbf{\ourmethod{} (Ours)}         & $\textbf{98.50}^*~(\small 0.0)$  & $\textbf{98.82}^*~(\small 0.0)$                                                   & $\textbf{87.76}^*~(\small 0.2)$  & $\textbf{89.22}^*~(\small 0.3)$
                               & $\textbf{79.25}^*~(\small 0.0)$  & $\textbf{56.62}~(\small 0.4)$    \\
        \bottomrule[1.5pt]
    \end{tabular}
    
    \label{tab:lp_results1}
\end{table*}

\subsection{Experiment Settings}
\subsubsection{Dataset}
We evaluate \ourmethod{} on three widely adopted Twitter bot detection benchmarks: Cresci-15 \cite{cresci2015fame}, TwiBot-20 \cite{feng2021twibot}, and TwiBot-22 \cite{feng2022twibot}. Cresci-15 \citep{cresci2015fame} is proposed in 2015 and includes 5,301 Twitter users and their network structure.
TwiBot-20 \citep{feng2021twibot} contains 229,580 users from the sports, economy, entertainment, and politics domains. TwiBot-22 \citep{feng2022twibot} is the most extensive dataset to date for Twitter bot detection, featuring a diverse set of entities and relationships on Twitter with 1,000,000 users. We follow the original train, valid, and test splits of the datasets for a fair comparison with previous works. 

\subsubsection{Baselines}
We compare \ourmethod{} with feature-, text-, and graph-based Twitter bot detection models:
\begin{itemize}[leftmargin=*]
    \item \textbf{SGBot} \cite{yang2020scalable} extracts features from users' metadata and feeds them into random forest classifiers for scalable and generalizable bot identification.
    \item \textbf{Wei \textit{et al.}} \cite{wei2019twitter} use word embeddings and bidirectional LSTMs to process user textual information for bot detection.
    \item \textbf{Varol \textit{et al.}} \cite{varol2017online} mainly use metadata, network, and other derived statistics to perform bot detection with random forest, aiming to detect different types of Twitter bots.
    \item \textbf{SATAR} \cite{feng2021satar} is a self-supervised approach for Twitter user representation learning, which use semantic, property, and neighborhood information and is further fine-tuned on Twitter bot detection datasets.
    \item \textbf{Botometer} \cite{yang2022botometer} use more than 1,000 features from user metadata, content, and interaction. They use an ensemble of existing datasets for training.
    \item \textbf{GraphHist} \cite{GraphHist} is a graph-based approach based on the latent feature histograms of users, which detects Twitter bots by users' ego graphs.
    \item \textbf{BotRGCN} \cite{feng2021botrgcn} builds a heterogeneous graph from the Twitter network and employs relational graph convolutional networks for user representation learning and Twitter bot detection.
    \item \textbf{RGT} \cite{feng2022heterogeneity} is short for relational graph transformers, which model the inherent heterogeneity in the Twittersphere to improve Twitter bot detection.
    \item \textbf{BotBuster} \cite{ng2022botbuster} is a social bot detection system that processes user metadata and textual information with mixture-of-experts to enhance cross-platform bot detection. 
    
\end{itemize}

\subsubsection{Implementation}
We use Pytorch \cite{pytorch2018pytorch}, Pytorch Geometric \cite{Fey/Lenssen/2019}, scikit-learn \cite{scikit-learn}, and Transformers \cite{wolf-etal-2020-transformers} to implement \ourmethod{}. The hyperparameter settings are presented in table \ref{tab:hyper} to facilitate reproduction. We conduct all experiments on a cluster with 4 Tesla V100 GPUs with 32 GB memory, 16 CPU cores, and 377GB CPU memory. It takes about 5 minutes, 1 hour, and 4 hours to train \ourmethod{} on the Cresci-15, TwiBot-20, and TwiBot-22. Code, data, and trained models will be made publicly available.

\begin{figure*}[ht]

    \centering
    \includegraphics[width=0.8\linewidth]{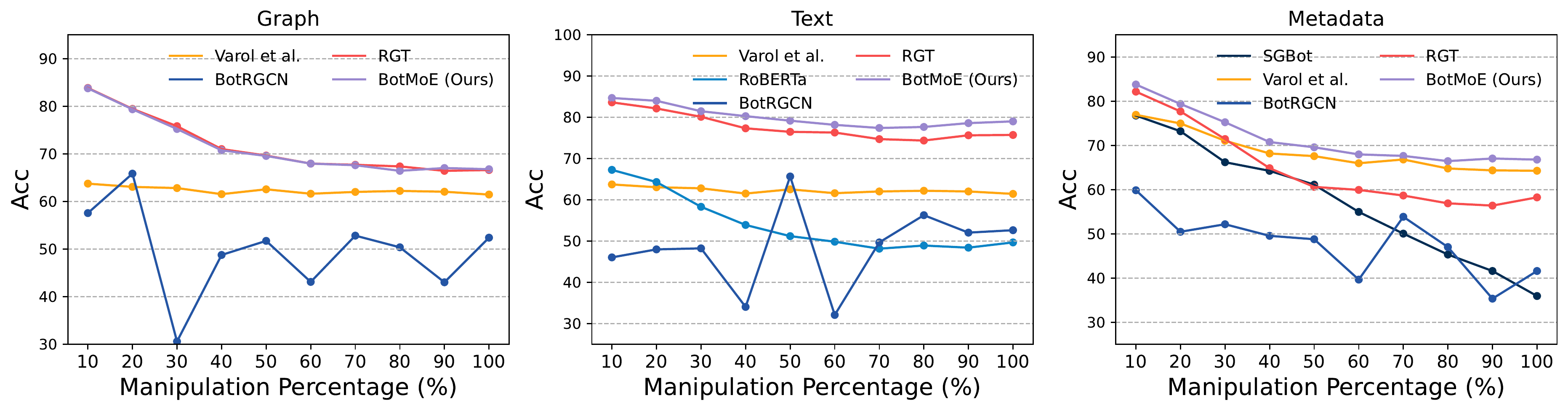}
    \caption{Twitter bot detection accuracy with different extent of graph, text, and metadata feature manipulation. \ourmethod{} consistently outperforms baseline models and shows a lower level of volatility.}
    \label{fig:manipulate}
\end{figure*} 

\begin{figure}
    \centering
    \includegraphics[width=0.6\linewidth]{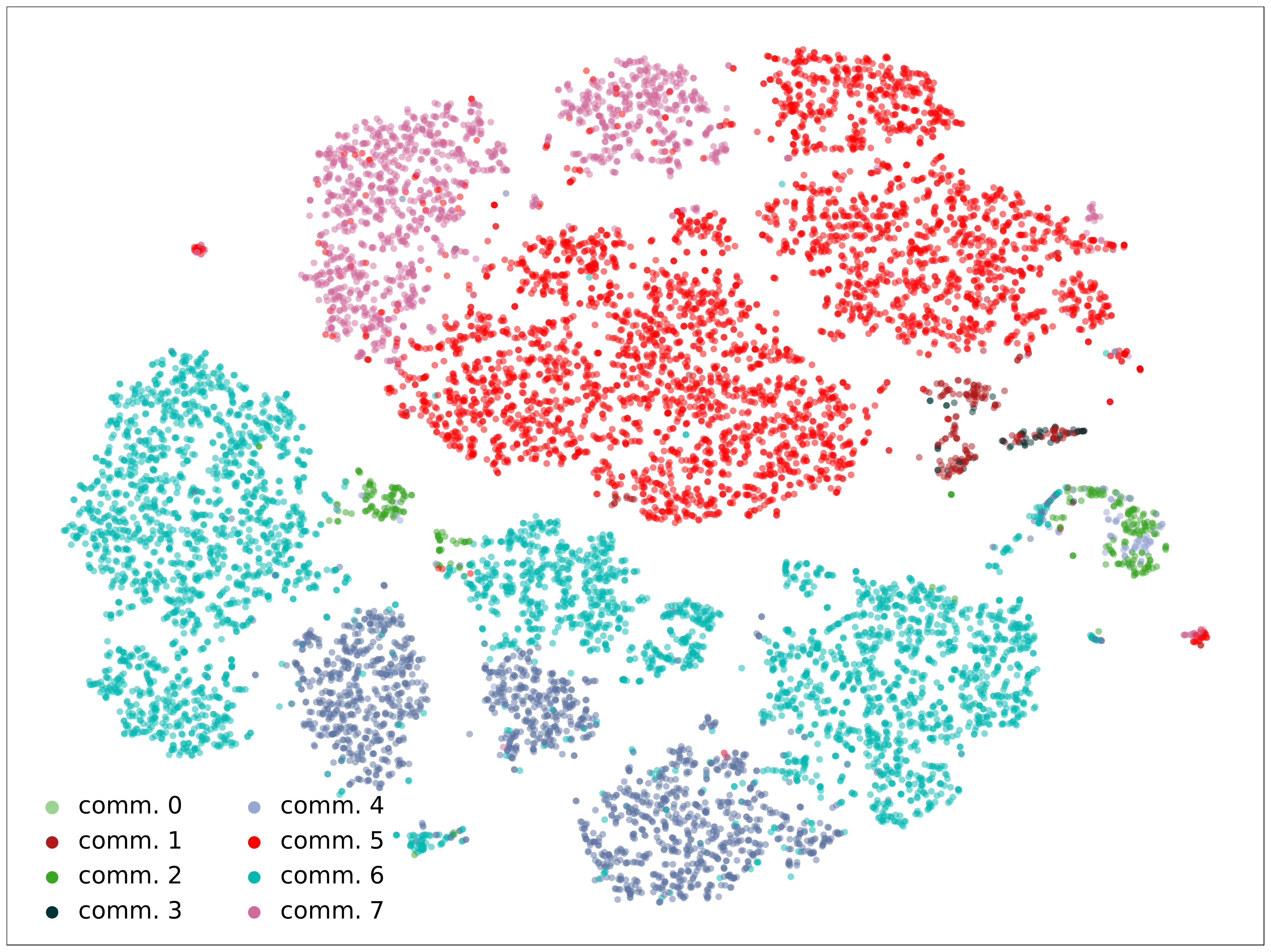}
    \caption{T-SNE visualization of users' graph, text, and metadata features, where users within the same communities are represented by the same color. The users are clearly divided into communities by MoE according to high-dimensional representation.}
    \label{fig:cluster}
\end{figure}

\subsection{Main Results}
\label{subsec:main_results}
We evaluate \ourmethod{} and 10 representative baselines on the three Twitter bot detection benchmarks and present the results in Table \ref{tab:lp_results1}, which demonstrates that:
\begin{itemize}[leftmargin=*]
    \item 
    \ourmethod{} consistently and significantly outperforms all baseline methods across the three datasets. Specifically, compared with the previously state-of-the-art RGT \cite{feng2022heterogeneity}, \ourmethod{} achieves \textbf{1.3\%} accuracy and \textbf{1.4\%} higher F1-score on TwiBot-20, and 2.0\% and 1.04\% F1-score on the TwiBot-22 benchmark, which are statistically significant improvements.

     \item Twitter bot detection models that leverage graphs and networks, such as RGT \citep{feng2022heterogeneity} and BotRGCN \citep{feng2021botrgcn}, generally outperform baselines that don't. This again reaffirms the importance of analyzing the network structure among Twitter users for robust Twitter bot detection. 
     
    \item Since these three datasets, from Cresci-15 to TwiBot-20 to TwiBot-22, present increasingly more up-to-date benchmarks with more advanced bots and user diversity, performance decline is generally witnessed across all bot detection models. Additionally, feature-based and text-based methods (SGBot and RoBERTa) tend to perform worse than graph-based methods (BotRGCN and RGT), due to the feature manipulation challenge most common among advanced bots \cite{cresci2020decade}. \ourmethod{} addresses this issue by using a multi-modal joint detection framework to comprehensively detect bots with forged features, resulting in the smallest performance decline on the more recent TwiBot-20 and TwiBot-22 benchmarks.

    \item Both BotBuster \cite{ng2022botbuster} and \ourmethod{} employ mixture-of-experts for Twitter bot detection. \ourmethod{} outperforms BotBuster on all three datasets, providing empirical evidence that our community-centric way of leveraging MoE is more robust.
\end{itemize}

\begin{figure*}[ht]

    \centering
    \includegraphics[width=0.8\linewidth]{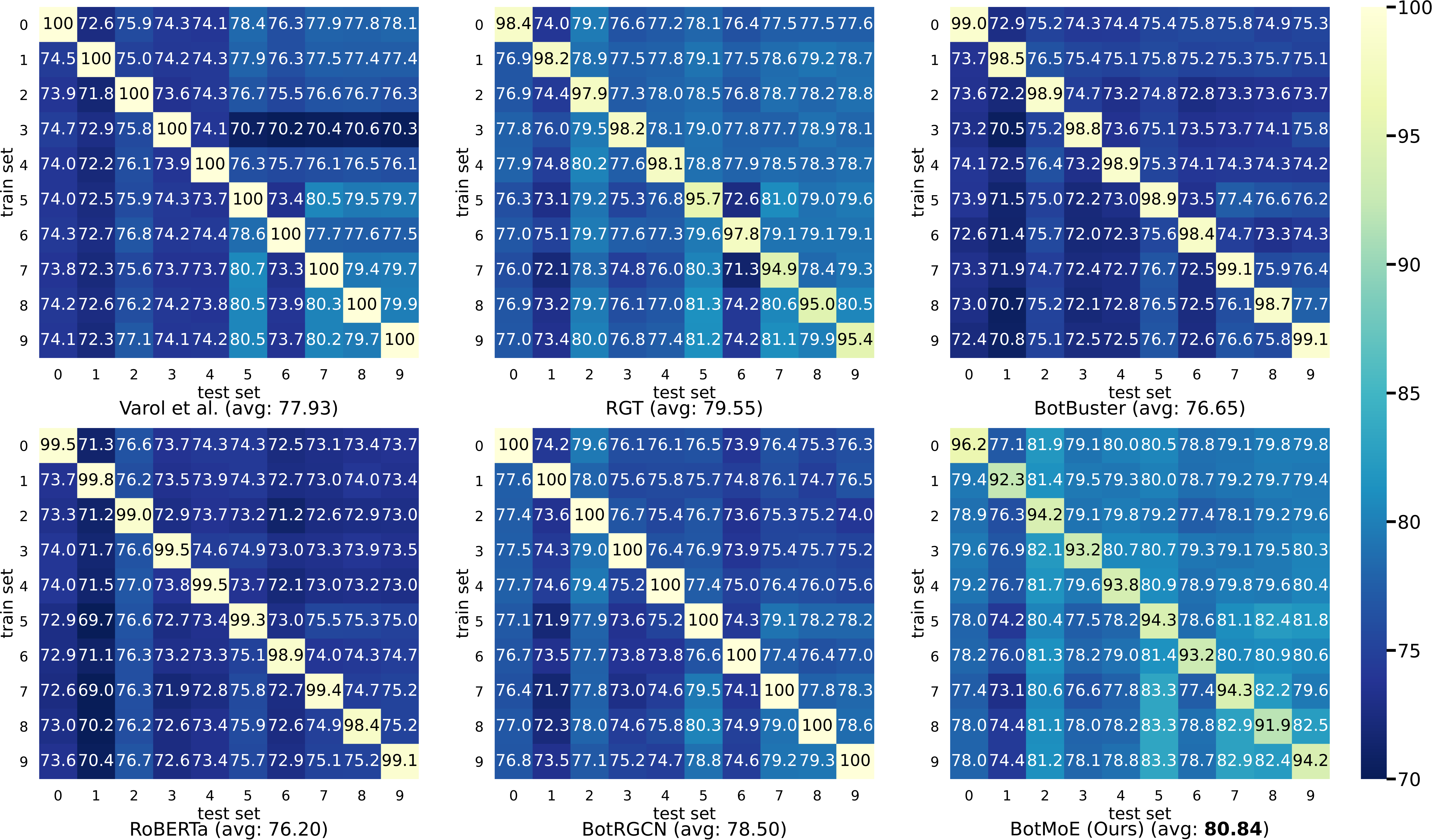}
    \caption{To test \ourmethod{}'s ability to generalize on unseen communities, we train and test on 10 sub-communities selected from Twibot-22 and report the accuracy. \ourmethod{} achieves the highest average accuracy, indicating better generalization. }
    \label{fig:generalization}
\end{figure*}

\subsection{Feature Manipulation Study}
To test the model's robustness over manipulated features, we amplify the extent of manipulation in graph, text, and metadata by replacing the features of bots with those of real users in test spit on TwiBot-20.
We used models trained on the original datasets to perform detection on the new test split and  plot their performance along with the manipulation. Specifically, for text and metadata, we randomly select human features to replace $10\%$-$100\%$ bot features. For graph information, based on the existing bot-human edges, we randomly add human-bot edges to increase the number by $10\%$-$100\%$.

As demonstrated in Figure \ref{fig:manipulate}, along with the increase of manipulated features, the decline of the performance of \ourmethod{} and RGT \cite{feng2022heterogeneity} is the least: \ourmethod{} has a decrease of only $9.9\%$ in textual information and only $23.9\%$ in $100\%$ manipulation in graph and metadata. In contrast, the simple text-based method RoBERTa reaches an accuracy of less than $50\%$ under the condition of $100\%$ deepfake texts. Besides, traditional feature-based machine learning methods SGBot and Varol don't perform well either: SGBot plummets from $76.75$ to $35.93$ under manipulated metadata feature.

\subsection{User Community Study}
\label{subsec:user_community_study}
To address the diverse community challenge, \ourmethod{} utilizes a gating network $G(\cdot)$ , according to equation \ref{eq:1}, to assign Twitter users to their corresponding communities and process them with community-specific expert networks. The accurate assignment of user communities is crucial for ensuring that the distributional difference across user communities is taken into account, thus we visualize the community assignments in \ourmethod{} in Figure \ref{fig:cluster}. Specifically, we extract user representations from the output of the MoE layer from the graph, text, and metadata modality and visualize them with T-SNE \cite{van2008visualizing}. Users with the same color are assigned to the same community and thus fed into the same expert network. Figure \ref{fig:cluster} demonstrates that users are clearly assigned under metadata, text, and graph modality, indicating the community-aware MoE layer's effectiveness in dividing users based on user representations.

\subsection{Generalization Study}
\label{subsec:generalization_study}
As Twitter bots are constantly evolving \cite{cresci2020decade}, the arms race between bot operators and bot detection research calls for models to better generalize to unseen accounts and user communities. To this end, we evaluate \ourmethod{}'s generalization ability to detect bots in unseen communities using the 10 sub-communities in the TwiBot-22 \citep{feng2022twibot} benchmark. As is presented in Figure \ref{fig:generalization}, we view these user communities as different data folds and examine the performance of \ourmethod{} and baseline models when trained on $i$-th fold and tested on $j$-th fold. It is illustrated that:

\begin{itemize}[leftmargin=*]
\item \textbf{\ourmethod{} is better at generalizing to unseen communities.} \ourmethod{} achieves the highest average accuracy of 80.84 among all bot detection approaches, which is 1.29\% higher than the state-of-the-art RGT \citep{feng2022heterogeneity}. This suggests that equipped with multi-modal user information analysis and the community-aware MoE layer, \ourmethod{} could better generalize to previously unseen accounts and user communities.
\item \textbf{Multi-modal methods better generalize than models with limited user features.} \ourmethod{} shows stronger generalization ability than feature-based, text-based, and graph-based models represented by \citet{varol2017online}, RoBERTa \cite{liu2019roberta}, and RGT \cite{feng2022heterogeneity}, respectively. This demonstrates that it is integral to leverage user information from multiple sources to enhance the model's generalization ability to unseen users.
\end{itemize}

\begin{figure}[t]
    \centering
    \includegraphics[width=0.8\linewidth]{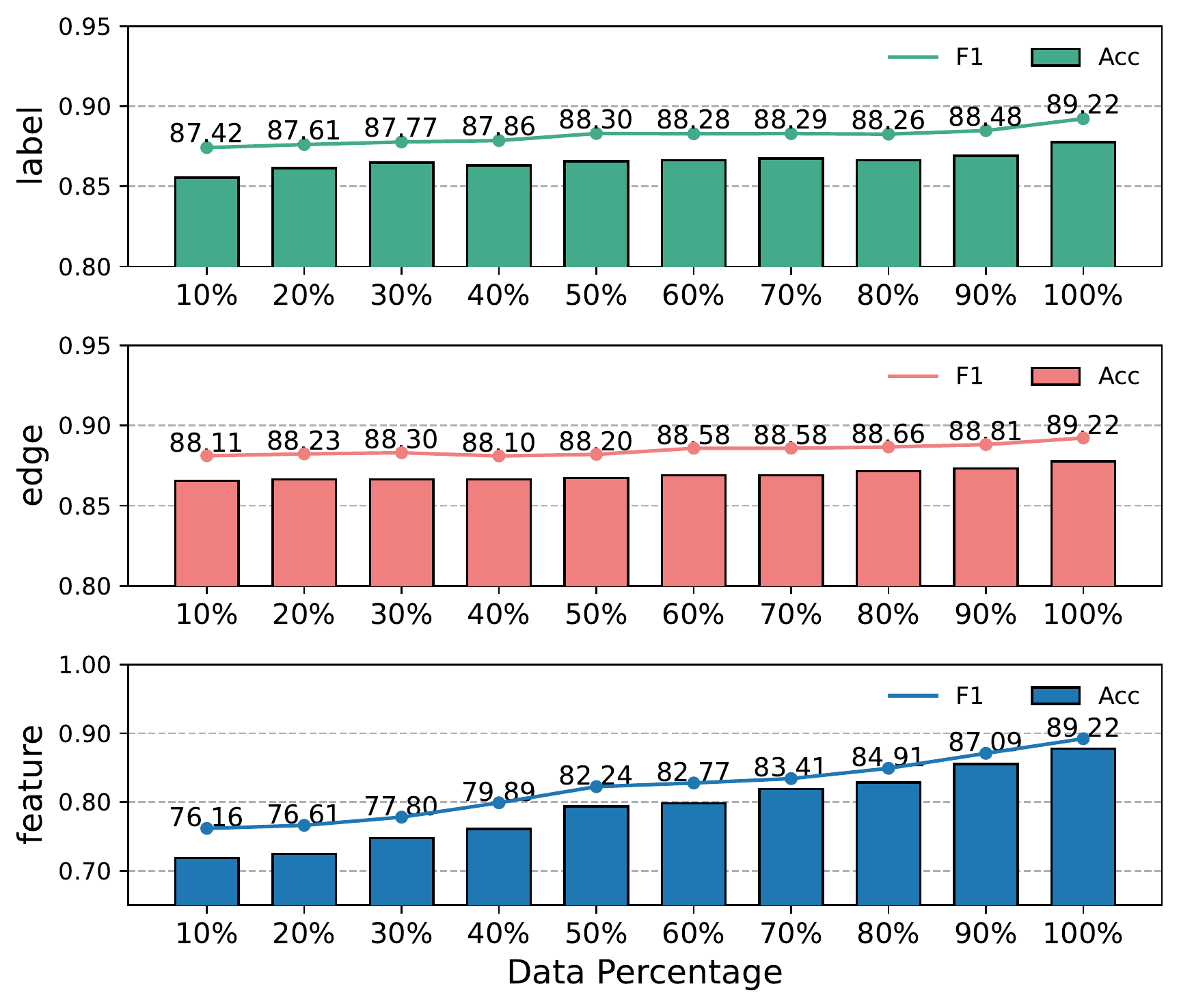}
    \caption{To test the robustness of our model, we randomly select $10\%$-$100\%$ of data on TwiBot-20 for training and test the model on the original test split and present the accuracy and F1-score. For edges or labels, it is demonstrated the drop of the F1-score remains within 2\% from 100\% data to only 10\% data provided, demonstrating the robustness of BoMoE against lack of data.} 
    \label{fig:data_efficiency}
\end{figure}

\begin{figure*}[ht]
    \centering
    \includegraphics[width=1\linewidth]{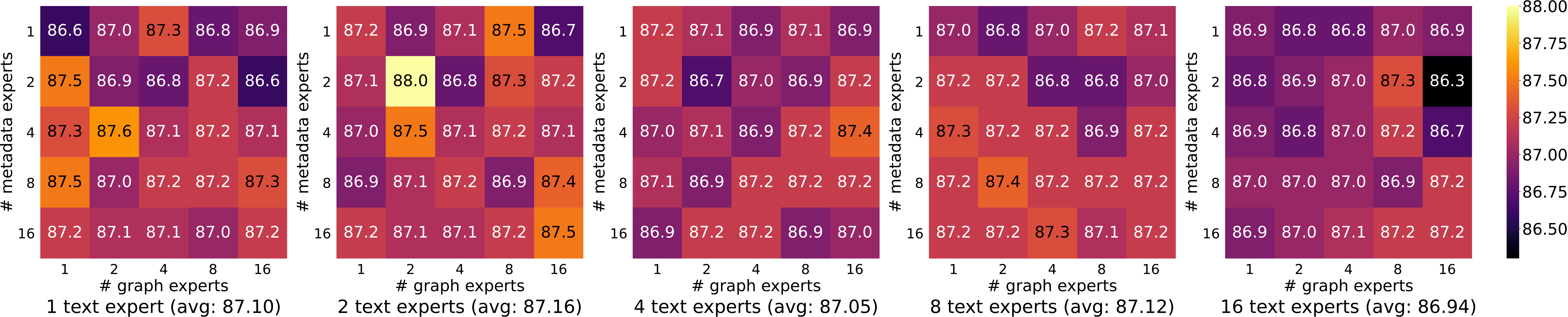}
    \caption{Accuracy score when increasing the number of experts of text, graph, and metadata. The best performance is achieved with 2 experts for each modality, proving that more experts don't necessarily promise high performance.}
    \label{fig:expert_num}
\end{figure*}

\subsection{Data Efficiency Study}
\label{subsec:efficiency_study}
Existing Twitter bot detection models over-rely on training data, requiring large quantities of labeled users for training and becoming insufficiently trained if otherwise. However, due to the obscure nature of advanced Twitter bots, obtaining accurate and reliable labels for Twitter bot detection is difficult, expensive, and noisy, creating challenges for the creation of Twitter bot detection datasets \cite{hays2023simplistic}. Therefore, it is essential to investigate \ourmethod{} and existing baselines' robustness against limited labels, user interactions, and noisy user features, which are common in existing bot detection datasets.
Specifically, we create ablated settings with partial training sets, randomly removed user interaction edges, and masked user features to re-evaluate bot detection models and present performance in Figure \ref{fig:data_efficiency}.
It demonstrates that even with only 60\% of user interactions and annotations, \ourmethod{} outperforms the full version of the state-of-the-art model RGT \cite{feng2022heterogeneity}, proving that \ourmethod{} is more robust towards fewer training data. 
We also find that when the number of user features is reduced, \ourmethod{} exhibits performance drops. We suspect this limitation could be in part alleviated by employing fewer experts in the community-aware MoE layers, which we further discuss in Section \ref{subsec:expert_num_study}.

\begin{table}
    \centering
    \caption{Ablation study of \ourmethod{}. \textbf{Bold} indicates the best performance, \underline{underline} the second best, and $"^*"$ denotes that the results are significantly better than the second-best under the student t-test.}
    \resizebox{1\linewidth}{!}{
    \begin{tabular}{c|c|cc} 
    \toprule[1.5pt]
    \textbf{Category}
    &{ \textbf{Ablation Settings}}      & $\textbf{Accuracy}^*$  & $\textbf{F1-score}^{*}$ \\  
     \midrule[0.75pt]
    \multirow{1}{*}{full model}&
    \ourmethod{}             & $\textbf{87.76}~(\small 0.2)$    & $\textbf{89.22}~(\small 0.3)$ \\
    \midrule[0.75pt]
    \multirow{3}{*}{community-aware}&
     replace MoE with MLP       & $86.90~(\small 0.0)$ & $88.44~(\small 0.0)$  \\
     &fully activated MoE       & $86.91~(\small 0.0)$ & $88.62~(\small 0.0)$  \\ 
     &w/o MoE                   & $86.88~(\small 0.0)$ & $88.41~(\small 0.0)$  \\ 
    \midrule[0.75pt]
    \multirow{3}{*}{modality}
    &w/o graph           & $80.55~(\small 0.1)$ & $82.47~(\small 0.2)$           \\
    &w/o text            & $86.95~(\small 0.1)$ & $88.10~(\small 0.7)$          \\
    &w/o metadata        & $87.10~(\small 0.0)$ & $88.67~(\small 0.0)$          \\
    \midrule[0.75pt]
    \multirow{4}{*}{modality fusion}
    & mean pooling                 & $\underline{87.37}~(\small 0.0)$ & $\underline{88.92}~(\small 0.0)$ \\
    & max pooling                   & $87.03~(\small 0.0)$ & $88.51~(\small 0.0)$ \\
    & min pooling                   & $87.23~(\small 0.0)$ & $88.73~(\small 0.0)$ \\
    & MLP                           & $87.07~(\small 0.0)$ & $88.59~(\small 0.0)$ \\
    \bottomrule[1.5pt]
    \end{tabular}
    }
    
    \label{tab:ablation}
\end{table}
\subsection{Ablation Study}
\label{subsec:ablation_study}
As \ourmethod{} outperforms baselines across all three benchmark datasets, we conduct further analysis to examine the impact of our original proposals on model performance with the Twibot-20 dataset and present the results in Table \ref{tab:ablation}:
\begin{itemize}[leftmargin=*]
    \item To study the impact of the community-aware mixture-of-experts layer, we replace it with a singular expert network, an MLP, and a fully activated MoE layer, or remove the community-aware layer. By replacing MoE with an expert, the accuracy decreases the least, only $0.7\%$ compared to full model. However, when the community-aware layer is fully removed, the model achieves the lowest performance presented in this table, only $86.88$ in accuracy, demonstrating the importance of incorporating a community-aware module that can adapt to diverse user communities and thus tackle the diverse community challenge. The MLP setting or the fully activated setting doesn't outperform the full model either. This indicates that the accuracy increase brought by MoE is not simply from enlarging the model's capacity.
    
    \item To evaluate the effectiveness of leveraging multiple user information modalities, we remove the metadata, text, and graph encoders and compare the ablated version to the full model. The modality category in Table \ref{tab:ablation} reveals that removing any modality would lead to performance drops, while the graph encoder is most critical to model performance, resulting in an $8.0\%$ drop in model accuracy. As a result, while the network structure of Twitter communities is of the most value, all three modalities are beneficial to \ourmethod{} and Twitter bot detection.
    
    \item To investigate the importance of our proposed expert fusion layer, we replace it with four simple aggregation strategies: mean, max, min, and MLP. Results in the fusion category in Table \ref{tab:ablation} demonstrate that the full \ourmethod{} outperforms all ablated models, validating the effectiveness of the expert fusion layer that enables information exchange across modalities. 
\end{itemize}


\subsection{Increasing the number of experts}
\label{subsec:expert_num_study}

An important parameter in the \ourmethod{} framework is the number of experts in each modality, which governs the number of communities and the inductive bias of \ourmethod{}. To investigate the optimal setup of experts, we conduct grid search and present the model's performance on Twibot-20 in Figure \ref{fig:expert_num}.
Overall, model performance is worst for the non-MoE setting (1 expert only). However, more experts don't necessarily guarantee better performance: the highest performance is achieved at $2$ text experts, $2$ graph experts, and $2$ metadata experts. This is because compared to the other large language models trained on colossal datasets, BotMoE is rather a lightweight model trained on a small part of social information. The total parameters of our model are less than 1 million, contrasting to large language models with millions of parameters in \citet{shazeer2017outrageously}. Therefore, the increase in the capacity may lead the model to overfit the training data. Besides, another finding is that models with more experts tend to be unstable during training perhaps because of the special nature of MoE.

\subsection{Case Study}
\label{subsec:case_study}

\begin{figure}[t]
	\centering
	\includegraphics[width=1\linewidth]{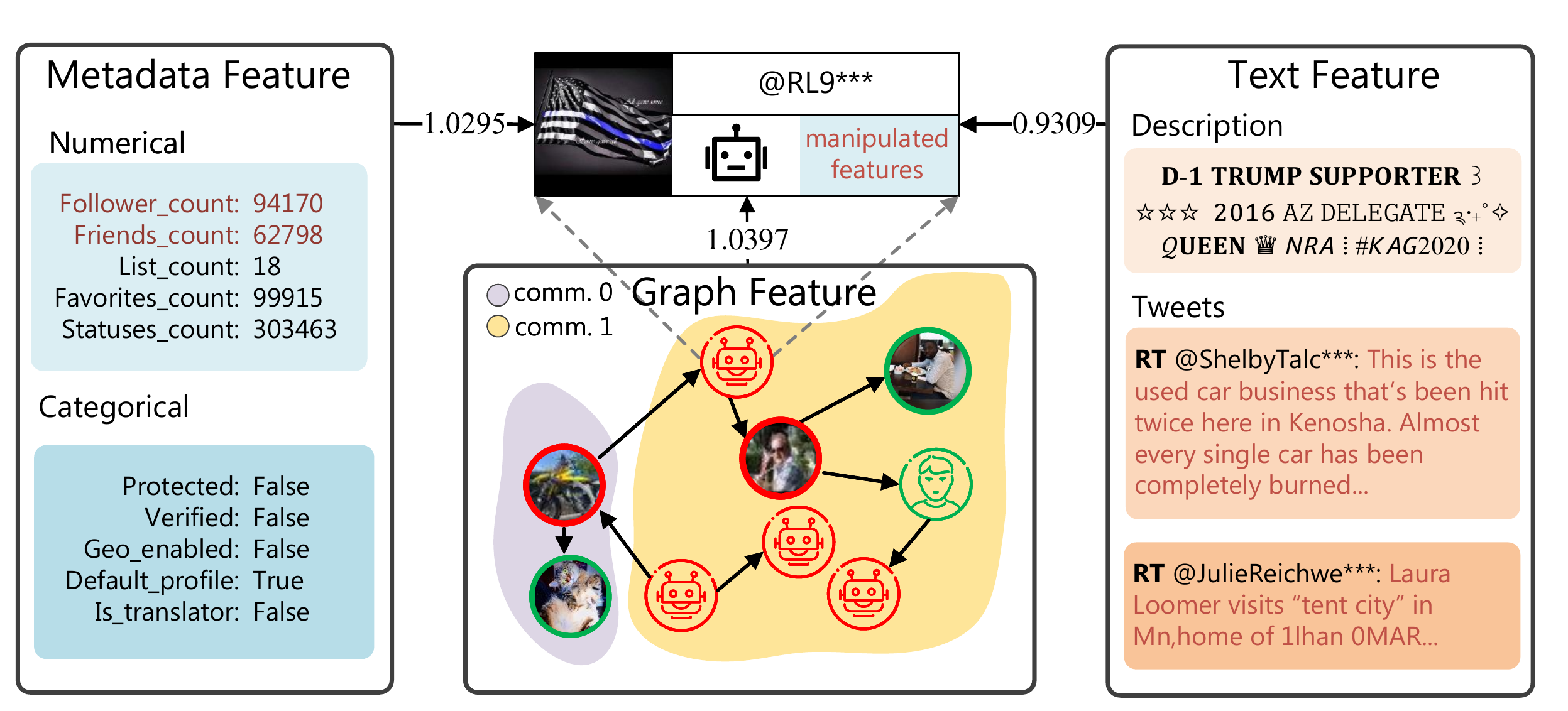}
         \caption{Case study: we display an advanced bot account with its graph, text, and metadata information and the corresponding consistency scores for outputs in \ourmethod{}. In the condition that the account metadata and the text information are manipulated and not highly distinguishable, \ourmethod{} weighs the graph structure as the most important and captures the clue of the bot account from local communities.
    	}
	\label{fig:case_study}
\end{figure}

We study a specific advanced Twitter bot and its attention weights from the expert fusion layer of \ourmethod{}. The scores are calculated from the sum of the attention weights from the multi-head attention layer to indicate the contribution of the feature of a specific type: graph, text, or user metadata. 

As illustrated in Figure \ref{fig:case_study}, the graph context demonstrates that the user's neighbors are mostly bot accounts, increasing the probability of bot prediction. Additionally, we observe that users with the same label and denser connections on the graph are more likely to be grouped within the same community, indicating the explainability of the community-aware MoE layer.

It is noteworthy that the user's tweets are highly misleading and may result in false positives when using bot detection methods. 
At first glance, the bot's recent retweets appear normal and it is challenging to distinguish this account from that of a genuine user. However, upon closer examination of the bot's metadata and graph features, the excessive status count and many bot neighbors indicate it is not a genuine user. From the attention weights of the expert fusion layer, we find metadata and graph features carry higher weights and provide more trustworthy information for bot detection. Therefore, relying solely on one modality of features for prediction may not be effective, as the model would not be able to utilize all the information that bots produce in order to identify them. However, \ourmethod{}'s multi-modal joint detection framework enables the model to focus on the more informative features and provide more reliable results.



\section{Discussion}
\paragraph{Limitations}
\ourmethod{} has several limitations:
First, more modalities could be introduced in \ourmethod{}. Twitter content is not limited to textual tweets, as many accounts also post pictures and videos. Therefore, more user information and modalities could be introduced for Twitter bot detection.
Second, our study is limited to Twitter users due to the abundant resources on Twitter bot detection, while other social media platforms are underexplored and whether \ourmethod{} could generalize to Reddit or Facebook remains unclear.
Finally, the interpretability of \ourmethod{} is limited, as it lacks direct evidence of which modality and what information in that modality contributes most to the model's performance and a detailed explanation needs to be explored for how the communities are formed and identified. We leave it to future work on how to incorporate more user information modalities, evaluate \ourmethod{} on social media other than Twitter, and improve the interpretability of \ourmethod{} and Twitter bot detection in general.
\paragraph{Ethical considerations}

\ourmethod{} should serve as an aid rather than the sole decision maker in detecting Twitter bots, and human oversight is necessary for making final judgments when deployed on real-world social networks. First, \ourmethod{} can have false positive predictions, where legitimate accounts are classified as bots by mistake, which can have serious consequences for businesses or individuals who spread important information \citep{thieltges2016devil}. Second, as a combination of graph neural networks and language models, \ourmethod{} inherits the bias presented in these models. For example, language models could introduce stereotypes and social bias \cite{nadeem-etal-2021-stereoset, liang2021towards} and graph neural networks could lead to discrimination towards certain demographic groups \citep{dong2022edits}.

\section{Conclusion}
We propose \ourmethod{}, a novel Twitter bot detection framework with community-aware mixtures of modal-specific experts, aiming to tackle the challenges of malicious feature manipulation and diverse user communities. Extensive experiments demonstrate that \ourmethod{} significantly advances the state-of-the-art on three Twitter bot detection benchmarks. Further studies demonstrate the effectiveness of our original technical contributions, highlighting \ourmethod{}'s ability to tackle malicious feature manipulation and diverse Twitter communities, and proving the superior generalization ability and data efficiency of \ourmethod{}.

\section*{Acknowledgment}
This work was supported by the National Key Research and Development Program of China (No. 2022YFB3102600), National Nature Science Foundation of China (No. 62192781, No. 62272374, No. 62202367, No. 62250009, No. 62137002, No. 61937001), Innovative Research Group of the National Natural Science Foundation of China (61721002), Innovation Research Team of Ministry of Education (IRT\_17R86), Project of China Knowledge Center for Engineering Science and Technology, and Project of Chinese academy of engineering ``The Online and Offline Mixed Educational Service System for `The Belt and Road' Training in MOOC China''. We would like to express our gratitude for the support of K. C. Wong Education Foundation. We also appreciate the reviewers and chairs for their constructive feedback. Lastly, we would like to thank all LUD lab members for fostering a collaborative research environment.

\bibliographystyle{ACM-Reference-Format}
\bibliography{sample-sigconf}


\begin{thebibliography}{73}


\ifx \showCODEN    \undefined \def \showCODEN     #1{\unskip}     \fi
\ifx \showDOI      \undefined \def \showDOI       #1{#1}\fi
\ifx \showISBNx    \undefined \def \showISBNx     #1{\unskip}     \fi
\ifx \showISBNxiii \undefined \def \showISBNxiii  #1{\unskip}     \fi
\ifx \showISSN     \undefined \def \showISSN      #1{\unskip}     \fi
\ifx \showLCCN     \undefined \def \showLCCN      #1{\unskip}     \fi
\ifx \shownote     \undefined \def \shownote      #1{#1}          \fi
\ifx \showarticletitle \undefined \def \showarticletitle #1{#1}   \fi
\ifx \showURL      \undefined \def \showURL       {\relax}        \fi
\providecommand\bibfield[2]{#2}
\providecommand\bibinfo[2]{#2}
\providecommand\natexlab[1]{#1}
\providecommand\showeprint[2][]{arXiv:#2}

\bibitem[Ali~Alhosseini et~al\mbox{.}(2019)]%
        {ali2019detect}
\bibfield{author}{\bibinfo{person}{Seyed Ali~Alhosseini}, \bibinfo{person}{Raad
  Bin~Tareaf}, \bibinfo{person}{Pejman Najafi}, {and}
  \bibinfo{person}{Christoph Meinel}.} \bibinfo{year}{2019}\natexlab{}.
\newblock \showarticletitle{Detect me if you can: Spam bot detection using
  inductive representation learning}. In \bibinfo{booktitle}{\emph{Companion
  Proceedings of The 2019 World Wide Web Conference}}.
  \bibinfo{pages}{148--153}.
\newblock


\bibitem[Alothali et~al\mbox{.}(2023)]%
        {alothali2023sebd}
\bibfield{author}{\bibinfo{person}{Eiman Alothali}, \bibinfo{person}{Kadhim
  Hayawi}, {and} \bibinfo{person}{Hany Alashwal}.}
  \bibinfo{year}{2023}\natexlab{}.
\newblock \showarticletitle{SEBD: A Stream Evolving Bot Detection Framework
  with Application of PAC Learning Approach to Maintain Accuracy and Confidence
  Levels}.
\newblock \bibinfo{journal}{\emph{Applied Sciences}} \bibinfo{volume}{13},
  \bibinfo{number}{7} (\bibinfo{year}{2023}), \bibinfo{pages}{4443}.
\newblock


\bibitem[Bao et~al\mbox{.}(2021)]%
        {bao2021vlmo}
\bibfield{author}{\bibinfo{person}{Hangbo Bao}, \bibinfo{person}{Wenhui Wang},
  \bibinfo{person}{Li Dong}, \bibinfo{person}{Qiang Liu},
  \bibinfo{person}{Owais~Khan Mohammed}, \bibinfo{person}{Kriti Aggarwal},
  \bibinfo{person}{Subhojit Som}, {and} \bibinfo{person}{Furu Wei}.}
  \bibinfo{year}{2021}\natexlab{}.
\newblock \showarticletitle{Vlmo: Unified vision-language pre-training with
  mixture-of-modality-experts}.
\newblock \bibinfo{journal}{\emph{arXiv preprint arXiv:2111.02358}}
  (\bibinfo{year}{2021}).
\newblock


\bibitem[Brachten et~al\mbox{.}(2017)]%
        {brachten2017strategies}
\bibfield{author}{\bibinfo{person}{Florian Brachten}, \bibinfo{person}{Stefan
  Stieglitz}, \bibinfo{person}{Lennart Hofeditz}, \bibinfo{person}{Katharina
  Kloppenborg}, {and} \bibinfo{person}{Annette Reimann}.}
  \bibinfo{year}{2017}\natexlab{}.
\newblock \showarticletitle{Strategies and Influence of Social Bots in a 2017
  German state election-A case study on Twitter}.
\newblock \bibinfo{journal}{\emph{arXiv preprint arXiv:1710.07562}}
  (\bibinfo{year}{2017}).
\newblock


\bibitem[Chavoshi et~al\mbox{.}(2016)]%
        {chavoshi2016debot}
\bibfield{author}{\bibinfo{person}{Nikan Chavoshi}, \bibinfo{person}{Hossein
  Hamooni}, {and} \bibinfo{person}{Abdullah Mueen}.}
  \bibinfo{year}{2016}\natexlab{}.
\newblock \showarticletitle{Debot: Twitter bot detection via warped
  correlation.}. In \bibinfo{booktitle}{\emph{Icdm}},
  Vol.~\bibinfo{volume}{18}. \bibinfo{pages}{28--65}.
\newblock


\bibitem[Cresci(2020)]%
        {cresci2020decade}
\bibfield{author}{\bibinfo{person}{Stefano Cresci}.}
  \bibinfo{year}{2020}\natexlab{}.
\newblock \showarticletitle{A decade of social bot detection}.
\newblock \bibinfo{journal}{\emph{Commun. ACM}} \bibinfo{volume}{63},
  \bibinfo{number}{10} (\bibinfo{year}{2020}), \bibinfo{pages}{72--83}.
\newblock


\bibitem[Cresci et~al\mbox{.}(2015)]%
        {cresci2015fame}
\bibfield{author}{\bibinfo{person}{Stefano Cresci}, \bibinfo{person}{Roberto
  Di~Pietro}, \bibinfo{person}{Marinella Petrocchi}, \bibinfo{person}{Angelo
  Spognardi}, {and} \bibinfo{person}{Maurizio Tesconi}.}
  \bibinfo{year}{2015}\natexlab{}.
\newblock \showarticletitle{Fame for sale: Efficient detection of fake Twitter
  followers}.
\newblock \bibinfo{journal}{\emph{Decision Support Systems}}
  \bibinfo{volume}{80} (\bibinfo{year}{2015}), \bibinfo{pages}{56--71}.
\newblock


\bibitem[Cresci et~al\mbox{.}(2017)]%
        {cresci2017paradigm}
\bibfield{author}{\bibinfo{person}{Stefano Cresci}, \bibinfo{person}{Roberto
  Di~Pietro}, \bibinfo{person}{Marinella Petrocchi}, \bibinfo{person}{Angelo
  Spognardi}, {and} \bibinfo{person}{Maurizio Tesconi}.}
  \bibinfo{year}{2017}\natexlab{}.
\newblock \showarticletitle{The paradigm-shift of social spambots: Evidence,
  theories, and tools for the arms race}. In
  \bibinfo{booktitle}{\emph{Proceedings of the 26th international conference on
  world wide web companion}}. \bibinfo{pages}{963--972}.
\newblock


\bibitem[Cresci et~al\mbox{.}(2023)]%
        {cresci2023demystifying}
\bibfield{author}{\bibinfo{person}{Stefano Cresci}, \bibinfo{person}{Roberto
  Di~Pietro}, \bibinfo{person}{Angelo Spognardi}, \bibinfo{person}{Maurizio
  Tesconi}, {and} \bibinfo{person}{Marinella Petrocchi}.}
  \bibinfo{year}{2023}\natexlab{}.
\newblock \showarticletitle{Demystifying Misconceptions in Social Bots
  Research}.
\newblock \bibinfo{journal}{\emph{arXiv preprint arXiv:2303.17251}}
  (\bibinfo{year}{2023}).
\newblock


\bibitem[Davoudi et~al\mbox{.}(2020)]%
        {davoudi2020towards}
\bibfield{author}{\bibinfo{person}{Anahita Davoudi}, \bibinfo{person}{Ari~Z
  Klein}, \bibinfo{person}{Abeed Sarker}, {and} \bibinfo{person}{Graciela
  Gonzalez-Hernandez}.} \bibinfo{year}{2020}\natexlab{}.
\newblock \showarticletitle{Towards automatic bot detection in Twitter for
  health-related tasks}.
\newblock \bibinfo{journal}{\emph{AMIA Summits on Translational Science
  Proceedings}}  \bibinfo{volume}{2020} (\bibinfo{year}{2020}),
  \bibinfo{pages}{136}.
\newblock


\bibitem[Dehghan et~al\mbox{.}(2022)]%
        {dehghan2022detecting}
\bibfield{author}{\bibinfo{person}{Ashkan Dehghan}, \bibinfo{person}{Kinga
  Siuta}, \bibinfo{person}{Agata Skorupka}, \bibinfo{person}{Akshat Dubey},
  \bibinfo{person}{Andrei Betlen}, \bibinfo{person}{David Miller},
  \bibinfo{person}{Wei Xu}, \bibinfo{person}{Bogumil Kaminski}, {and}
  \bibinfo{person}{Pawel Pralat}.} \bibinfo{year}{2022}\natexlab{}.
\newblock \showarticletitle{Detecting Bots in Social-Networks Using Node and
  Structural Embeddings}.
\newblock  (\bibinfo{year}{2022}).
\newblock


\bibitem[Dong et~al\mbox{.}(2022)]%
        {dong2022edits}
\bibfield{author}{\bibinfo{person}{Yushun Dong}, \bibinfo{person}{Ninghao Liu},
  \bibinfo{person}{Brian Jalaian}, {and} \bibinfo{person}{Jundong Li}.}
  \bibinfo{year}{2022}\natexlab{}.
\newblock \showarticletitle{Edits: Modeling and mitigating data bias for graph
  neural networks}. In \bibinfo{booktitle}{\emph{Proceedings of the ACM Web
  Conference 2022}}. \bibinfo{pages}{1259--1269}.
\newblock


\bibitem[Duki{\'c} et~al\mbox{.}(2020)]%
        {dukic2020you}
\bibfield{author}{\bibinfo{person}{David Duki{\'c}}, \bibinfo{person}{Dominik
  Ke{\v{c}}a}, {and} \bibinfo{person}{Dominik Stipi{\'c}}.}
  \bibinfo{year}{2020}\natexlab{}.
\newblock \showarticletitle{Are you human? Detecting bots on Twitter Using
  BERT}. In \bibinfo{booktitle}{\emph{2020 IEEE 7th International Conference on
  Data Science and Advanced Analytics (DSAA)}}. IEEE,
  \bibinfo{pages}{631--636}.
\newblock


\bibitem[Everitt(1998)]%
        {everitt1998cambridge}
\bibfield{author}{\bibinfo{person}{B Everitt}.}
  \bibinfo{year}{1998}\natexlab{}.
\newblock \showarticletitle{The cambridge dictionary of statistics cambridge
  university press}.
\newblock \bibinfo{journal}{\emph{Cambridge, UK Google Scholar}}
  (\bibinfo{year}{1998}).
\newblock


\bibitem[Fedus et~al\mbox{.}(2021)]%
        {fedus2021switch}
\bibfield{author}{\bibinfo{person}{William Fedus}, \bibinfo{person}{Barret
  Zoph}, {and} \bibinfo{person}{Noam Shazeer}.}
  \bibinfo{year}{2021}\natexlab{}.
\newblock \bibinfo{title}{Switch transformers: Scaling to trillion parameter
  models with simple and efficient sparsity}.
\newblock
\newblock


\bibitem[Feng et~al\mbox{.}(2022a)]%
        {feng2022heterogeneity}
\bibfield{author}{\bibinfo{person}{Shangbin Feng}, \bibinfo{person}{Zhaoxuan
  Tan}, \bibinfo{person}{Rui Li}, {and} \bibinfo{person}{Minnan Luo}.}
  \bibinfo{year}{2022}\natexlab{a}.
\newblock \showarticletitle{Heterogeneity-aware twitter bot detection with
  relational graph transformers}. In \bibinfo{booktitle}{\emph{Proceedings of
  the AAAI Conference on Artificial Intelligence}}, Vol.~\bibinfo{volume}{36}.
  \bibinfo{pages}{3977--3985}.
\newblock


\bibitem[Feng et~al\mbox{.}(2022b)]%
        {feng2022twibot}
\bibfield{author}{\bibinfo{person}{Shangbin Feng}, \bibinfo{person}{Zhaoxuan
  Tan}, \bibinfo{person}{Herun Wan}, \bibinfo{person}{Ningnan Wang},
  \bibinfo{person}{Zilong Chen}, \bibinfo{person}{Binchi Zhang},
  \bibinfo{person}{Qinghua Zheng}, \bibinfo{person}{Wenqian Zhang},
  \bibinfo{person}{Zhenyu Lei}, \bibinfo{person}{Shujie Yang}, {et~al\mbox{.}}}
  \bibinfo{year}{2022}\natexlab{b}.
\newblock \showarticletitle{TwiBot-22: Towards Graph-Based Twitter Bot
  Detection}. In \bibinfo{booktitle}{\emph{Thirty-sixth Conference on Neural
  Information Processing Systems Datasets and Benchmarks Track}}.
\newblock


\bibitem[Feng et~al\mbox{.}(2021b)]%
        {feng2021satar}
\bibfield{author}{\bibinfo{person}{Shangbin Feng}, \bibinfo{person}{Herun Wan},
  \bibinfo{person}{Ningnan Wang}, \bibinfo{person}{Jundong Li}, {and}
  \bibinfo{person}{Minnan Luo}.} \bibinfo{year}{2021}\natexlab{b}.
\newblock \showarticletitle{Satar: A self-supervised approach to twitter
  account representation learning and its application in bot detection}. In
  \bibinfo{booktitle}{\emph{Proceedings of the 30th ACM International
  Conference on Information \& Knowledge Management}}.
  \bibinfo{pages}{3808--3817}.
\newblock


\bibitem[Feng et~al\mbox{.}(2021c)]%
        {feng2021twibot}
\bibfield{author}{\bibinfo{person}{Shangbin Feng}, \bibinfo{person}{Herun Wan},
  \bibinfo{person}{Ningnan Wang}, \bibinfo{person}{Jundong Li}, {and}
  \bibinfo{person}{Minnan Luo}.} \bibinfo{year}{2021}\natexlab{c}.
\newblock \showarticletitle{Twibot-20: A comprehensive twitter bot detection
  benchmark}. In \bibinfo{booktitle}{\emph{Proceedings of the 30th ACM
  International Conference on Information \& Knowledge Management}}.
  \bibinfo{pages}{4485--4494}.
\newblock


\bibitem[Feng et~al\mbox{.}(2021a)]%
        {feng2021botrgcn}
\bibfield{author}{\bibinfo{person}{Shangbin Feng}, \bibinfo{person}{Herun Wan},
  \bibinfo{person}{Ningnan Wang}, {and} \bibinfo{person}{Minnan Luo}.}
  \bibinfo{year}{2021}\natexlab{a}.
\newblock \showarticletitle{BotRGCN: Twitter bot detection with relational
  graph convolutional networks}. In \bibinfo{booktitle}{\emph{Proceedings of
  the 2021 IEEE/ACM International Conference on Advances in Social Networks
  Analysis and Mining}}. \bibinfo{pages}{236--239}.
\newblock


\bibitem[Ferrara(2017)]%
        {ferrara2017disinformation}
\bibfield{author}{\bibinfo{person}{Emilio Ferrara}.}
  \bibinfo{year}{2017}\natexlab{}.
\newblock \showarticletitle{Disinformation and social bot operations in the run
  up to the 2017 French presidential election}.
\newblock \bibinfo{journal}{\emph{arXiv preprint arXiv:1707.00086}}
  (\bibinfo{year}{2017}).
\newblock


\bibitem[Ferrara(2020)]%
        {ferrara2020covid}
\bibfield{author}{\bibinfo{person}{Emilio Ferrara}.}
  \bibinfo{year}{2020}\natexlab{}.
\newblock \showarticletitle{\# covid-19 on twitter: Bots, conspiracies, and
  social media activism}.
\newblock \bibinfo{journal}{\emph{arXiv preprint arXiv: 2004.09531}}
  (\bibinfo{year}{2020}).
\newblock


\bibitem[Ferrara(2022)]%
        {ferrara2022twitter}
\bibfield{author}{\bibinfo{person}{Emilio Ferrara}.}
  \bibinfo{year}{2022}\natexlab{}.
\newblock \showarticletitle{Twitter Spam and False Accounts Prevalence,
  Detection and Characterization: A Survey}.
\newblock \bibinfo{journal}{\emph{arXiv preprint arXiv:2211.05913}}
  (\bibinfo{year}{2022}).
\newblock


\bibitem[Fey and Lenssen(2019)]%
        {Fey/Lenssen/2019}
\bibfield{author}{\bibinfo{person}{Matthias Fey} {and} \bibinfo{person}{Jan~E.
  Lenssen}.} \bibinfo{year}{2019}\natexlab{}.
\newblock \showarticletitle{Fast Graph Representation Learning with {PyTorch
  Geometric}}. In \bibinfo{booktitle}{\emph{ICLR Workshop on Representation
  Learning on Graphs and Manifolds}}.
\newblock


\bibitem[Flores-Saviaga et~al\mbox{.}(2022)]%
        {flores2022datavoidant}
\bibfield{author}{\bibinfo{person}{Claudia Flores-Saviaga},
  \bibinfo{person}{Shangbin Feng}, {and} \bibinfo{person}{Saiph Savage}.}
  \bibinfo{year}{2022}\natexlab{}.
\newblock \showarticletitle{Datavoidant: An AI System for Addressing Political
  Data Voids on Social Media}.
\newblock \bibinfo{journal}{\emph{Proceedings of the ACM on Human-Computer
  Interaction}} \bibinfo{volume}{6}, \bibinfo{number}{CSCW2}
  (\bibinfo{year}{2022}), \bibinfo{pages}{1--29}.
\newblock


\bibitem[Greve et~al\mbox{.}(2022)]%
        {greve2022online}
\bibfield{author}{\bibinfo{person}{Henrich~R Greve},
  \bibinfo{person}{Hayagreeva Rao}, \bibinfo{person}{Paul Vicinanza}, {and}
  \bibinfo{person}{Echo~Yan Zhou}.} \bibinfo{year}{2022}\natexlab{}.
\newblock \showarticletitle{Online Conspiracy Groups: Micro-Bloggers, Bots, and
  Coronavirus Conspiracy Talk on Twitter}.
\newblock \bibinfo{journal}{\emph{American Sociological Review}}
  \bibinfo{volume}{87}, \bibinfo{number}{6} (\bibinfo{year}{2022}),
  \bibinfo{pages}{919--949}.
\newblock


\bibitem[Hayawi et~al\mbox{.}(2022)]%
        {hayawi2022deeprobot}
\bibfield{author}{\bibinfo{person}{Kadhim Hayawi}, \bibinfo{person}{Sujith
  Mathew}, \bibinfo{person}{Neethu Venugopal}, \bibinfo{person}{Mohammad~M
  Masud}, {and} \bibinfo{person}{Pin-Han Ho}.} \bibinfo{year}{2022}\natexlab{}.
\newblock \showarticletitle{DeeProBot: a hybrid deep neural network model for
  social bot detection based on user profile data}.
\newblock \bibinfo{journal}{\emph{Social Network Analysis and Mining}}
  \bibinfo{volume}{12}, \bibinfo{number}{1} (\bibinfo{year}{2022}),
  \bibinfo{pages}{1--19}.
\newblock


\bibitem[Hays et~al\mbox{.}(2023)]%
        {hays2023simplistic}
\bibfield{author}{\bibinfo{person}{Chris Hays}, \bibinfo{person}{Zachary
  Schutzman}, \bibinfo{person}{Manish Raghavan}, \bibinfo{person}{Erin Walk},
  {and} \bibinfo{person}{Philipp Zimmer}.} \bibinfo{year}{2023}\natexlab{}.
\newblock \showarticletitle{Simplistic Collection and Labeling Practices Limit
  the Utility of Benchmark Datasets for Twitter Bot Detection}.
\newblock \bibinfo{journal}{\emph{arXiv preprint arXiv:2301.07015}}
  (\bibinfo{year}{2023}).
\newblock


\bibitem[Heidari and Jones(2020)]%
        {heidari2020using}
\bibfield{author}{\bibinfo{person}{Maryam Heidari} {and}
  \bibinfo{person}{James~H Jones}.} \bibinfo{year}{2020}\natexlab{}.
\newblock \showarticletitle{Using bert to extract topic-independent sentiment
  features for social media bot detection}. In \bibinfo{booktitle}{\emph{2020
  11th IEEE Annual Ubiquitous Computing, Electronics \& Mobile Communication
  Conference (UEMCON)}}. IEEE, \bibinfo{pages}{0542--0547}.
\newblock


\bibitem[Hu et~al\mbox{.}(2021)]%
        {hu2021graph}
\bibfield{author}{\bibinfo{person}{Fenyu Hu}, \bibinfo{person}{Liping Wang},
  \bibinfo{person}{Shu Wu}, \bibinfo{person}{Liang Wang}, {and}
  \bibinfo{person}{Tieniu Tan}.} \bibinfo{year}{2021}\natexlab{}.
\newblock \showarticletitle{Graph classification by mixture of diverse
  experts}.
\newblock \bibinfo{journal}{\emph{arXiv preprint arXiv:2103.15622}}
  (\bibinfo{year}{2021}).
\newblock


\bibitem[Hu et~al\mbox{.}(2020)]%
        {hu2020heterogeneous}
\bibfield{author}{\bibinfo{person}{Ziniu Hu}, \bibinfo{person}{Yuxiao Dong},
  \bibinfo{person}{Kuansan Wang}, {and} \bibinfo{person}{Yizhou Sun}.}
  \bibinfo{year}{2020}\natexlab{}.
\newblock \showarticletitle{Heterogeneous graph transformer}. In
  \bibinfo{booktitle}{\emph{Proceedings of The Web Conference 2020}}.
  \bibinfo{pages}{2704--2710}.
\newblock


\bibitem[Kipf and Welling(2016)]%
        {kipf2016semi}
\bibfield{author}{\bibinfo{person}{Thomas~N Kipf} {and} \bibinfo{person}{Max
  Welling}.} \bibinfo{year}{2016}\natexlab{}.
\newblock \showarticletitle{Semi-supervised classification with graph
  convolutional networks}.
\newblock \bibinfo{journal}{\emph{arXiv preprint arXiv:1609.02907}}
  (\bibinfo{year}{2016}).
\newblock


\bibitem[Kudugunta and Ferrara(2018)]%
        {kudugunta2018deep}
\bibfield{author}{\bibinfo{person}{Sneha Kudugunta} {and}
  \bibinfo{person}{Emilio Ferrara}.} \bibinfo{year}{2018}\natexlab{}.
\newblock \showarticletitle{Deep neural networks for bot detection}.
\newblock \bibinfo{journal}{\emph{Information Sciences}}  \bibinfo{volume}{467}
  (\bibinfo{year}{2018}), \bibinfo{pages}{312--322}.
\newblock


\bibitem[Kumarage et~al\mbox{.}(2023)]%
        {kumarage2023stylometric}
\bibfield{author}{\bibinfo{person}{Tharindu Kumarage}, \bibinfo{person}{Joshua
  Garland}, \bibinfo{person}{Amrita Bhattacharjee}, \bibinfo{person}{Kirill
  Trapeznikov}, \bibinfo{person}{Scott Ruston}, {and} \bibinfo{person}{Huan
  Liu}.} \bibinfo{year}{2023}\natexlab{}.
\newblock \showarticletitle{Stylometric Detection of AI-Generated Text in
  Twitter Timelines}.
\newblock \bibinfo{journal}{\emph{arXiv preprint arXiv:2303.03697}}
  (\bibinfo{year}{2023}).
\newblock


\bibitem[Lei et~al\mbox{.}(2022)]%
        {lei2022bic}
\bibfield{author}{\bibinfo{person}{Zhenyu Lei}, \bibinfo{person}{Herun Wan},
  \bibinfo{person}{Wenqian Zhang}, \bibinfo{person}{Shangbin Feng},
  \bibinfo{person}{Zilong Chen}, \bibinfo{person}{Qinghua Zheng}, {and}
  \bibinfo{person}{Minnan Luo}.} \bibinfo{year}{2022}\natexlab{}.
\newblock \showarticletitle{BIC: Twitter Bot Detection with Text-Graph
  Interaction and Semantic Consistency}.
\newblock \bibinfo{journal}{\emph{arXiv preprint arXiv:2208.08320}}
  (\bibinfo{year}{2022}).
\newblock


\bibitem[Li et~al\mbox{.}(2022)]%
        {li2022botfinder}
\bibfield{author}{\bibinfo{person}{Shudong Li}, \bibinfo{person}{Chuanyu Zhao},
  \bibinfo{person}{Qing Li}, \bibinfo{person}{Jiuming Huang},
  \bibinfo{person}{Dawei Zhao}, {and} \bibinfo{person}{Peican Zhu}.}
  \bibinfo{year}{2022}\natexlab{}.
\newblock \showarticletitle{BotFinder: a novel framework for social bots
  detection in online social networks based on graph embedding and community
  detection}.
\newblock \bibinfo{journal}{\emph{World Wide Web}} (\bibinfo{year}{2022}),
  \bibinfo{pages}{1--17}.
\newblock


\bibitem[Liang et~al\mbox{.}(2021)]%
        {liang2021towards}
\bibfield{author}{\bibinfo{person}{Paul~Pu Liang}, \bibinfo{person}{Chiyu Wu},
  \bibinfo{person}{Louis-Philippe Morency}, {and} \bibinfo{person}{Ruslan
  Salakhutdinov}.} \bibinfo{year}{2021}\natexlab{}.
\newblock \showarticletitle{Towards understanding and mitigating social biases
  in language models}. In \bibinfo{booktitle}{\emph{ICML}}. PMLR,
  \bibinfo{pages}{6565--6576}.
\newblock


\bibitem[Liu et~al\mbox{.}(2019)]%
        {liu2019roberta}
\bibfield{author}{\bibinfo{person}{Yinhan Liu}, \bibinfo{person}{Myle Ott},
  \bibinfo{person}{Naman Goyal}, \bibinfo{person}{Jingfei Du},
  \bibinfo{person}{Mandar Joshi}, \bibinfo{person}{Danqi Chen},
  \bibinfo{person}{Omer Levy}, \bibinfo{person}{Mike Lewis},
  \bibinfo{person}{Luke Zettlemoyer}, {and} \bibinfo{person}{Veselin
  Stoyanov}.} \bibinfo{year}{2019}\natexlab{}.
\newblock \showarticletitle{Roberta: A robustly optimized bert pretraining
  approach}.
\newblock \bibinfo{journal}{\emph{arXiv preprint arXiv:1907.11692}}
  (\bibinfo{year}{2019}).
\newblock


\bibitem[Luo et~al\mbox{.}(2020)]%
        {luo2020deepbot}
\bibfield{author}{\bibinfo{person}{Linhao Luo}, \bibinfo{person}{Xiaofeng
  Zhang}, \bibinfo{person}{Xiaofei Yang}, {and} \bibinfo{person}{Weihuang
  Yang}.} \bibinfo{year}{2020}\natexlab{}.
\newblock \showarticletitle{Deepbot: a deep neural network based approach for
  detecting Twitter bots}. In \bibinfo{booktitle}{\emph{IOP Conference Series:
  Materials Science and Engineering}}, Vol.~\bibinfo{volume}{719}. IOP
  Publishing, \bibinfo{pages}{012063}.
\newblock


\bibitem[Ma et~al\mbox{.}(2018)]%
        {ma2018modeling}
\bibfield{author}{\bibinfo{person}{Jiaqi Ma}, \bibinfo{person}{Zhe Zhao},
  \bibinfo{person}{Xinyang Yi}, \bibinfo{person}{Jilin Chen},
  \bibinfo{person}{Lichan Hong}, {and} \bibinfo{person}{Ed~H Chi}.}
  \bibinfo{year}{2018}\natexlab{}.
\newblock \showarticletitle{Modeling task relationships in multi-task learning
  with multi-gate mixture-of-experts}. In \bibinfo{booktitle}{\emph{Proceedings
  of the 24th ACM SIGKDD international conference on knowledge discovery \&
  data mining}}. \bibinfo{pages}{1930--1939}.
\newblock


\bibitem[Madaan et~al\mbox{.}(2021)]%
        {madaan2021think}
\bibfield{author}{\bibinfo{person}{Aman Madaan}, \bibinfo{person}{Niket
  Tandon}, \bibinfo{person}{Dheeraj Rajagopal}, \bibinfo{person}{Peter Clark},
  \bibinfo{person}{Yiming Yang}, {and} \bibinfo{person}{Eduard Hovy}.}
  \bibinfo{year}{2021}\natexlab{}.
\newblock \showarticletitle{Think about it! Improving defeasible reasoning by
  first modeling the question scenario}.
\newblock \bibinfo{journal}{\emph{arXiv preprint arXiv:2110.12349}}
  (\bibinfo{year}{2021}).
\newblock


\bibitem[Magelinski et~al\mbox{.}(2020)]%
        {GraphHist}
\bibfield{author}{\bibinfo{person}{Thomas Magelinski}, \bibinfo{person}{David
  Beskow}, {and} \bibinfo{person}{Kathleen~M Carley}.}
  \bibinfo{year}{2020}\natexlab{}.
\newblock \showarticletitle{Graph-hist: Graph classification from latent
  feature histograms with application to bot detection}. In
  \bibinfo{booktitle}{\emph{Proceedings of the AAAI Conference on Artificial
  Intelligence}}, Vol.~\bibinfo{volume}{34}. \bibinfo{pages}{5134--5141}.
\newblock


\bibitem[Miller et~al\mbox{.}(2014)]%
        {miller2014twitter}
\bibfield{author}{\bibinfo{person}{Zachary Miller}, \bibinfo{person}{Brian
  Dickinson}, \bibinfo{person}{William Deitrick}, \bibinfo{person}{Wei Hu},
  {and} \bibinfo{person}{Alex~Hai Wang}.} \bibinfo{year}{2014}\natexlab{}.
\newblock \showarticletitle{Twitter spammer detection using data stream
  clustering}.
\newblock \bibinfo{journal}{\emph{Information Sciences}}  \bibinfo{volume}{260}
  (\bibinfo{year}{2014}), \bibinfo{pages}{64--73}.
\newblock


\bibitem[Morstatter et~al\mbox{.}(2016)]%
        {morstatter2016new}
\bibfield{author}{\bibinfo{person}{Fred Morstatter}, \bibinfo{person}{Liang
  Wu}, \bibinfo{person}{Tahora~H Nazer}, \bibinfo{person}{Kathleen~M Carley},
  {and} \bibinfo{person}{Huan Liu}.} \bibinfo{year}{2016}\natexlab{}.
\newblock \showarticletitle{A new approach to bot detection: striking the
  balance between precision and recall}. In \bibinfo{booktitle}{\emph{2016
  IEEE/ACM International Conference on Advances in Social Networks Analysis and
  Mining (ASONAM)}}. IEEE, \bibinfo{pages}{533--540}.
\newblock


\bibitem[Nadeem et~al\mbox{.}(2021)]%
        {nadeem-etal-2021-stereoset}
\bibfield{author}{\bibinfo{person}{Moin Nadeem}, \bibinfo{person}{Anna Bethke},
  {and} \bibinfo{person}{Siva Reddy}.} \bibinfo{year}{2021}\natexlab{}.
\newblock \showarticletitle{{S}tereo{S}et: Measuring stereotypical bias in
  pretrained language models}. In \bibinfo{booktitle}{\emph{Proceedings of the
  59th Annual Meeting of the Association for Computational Linguistics and the
  11th International Joint Conference on Natural Language Processing (Volume 1:
  Long Papers)}}. \bibinfo{publisher}{Association for Computational
  Linguistics}, \bibinfo{address}{Online}, \bibinfo{pages}{5356--5371}.
\newblock
\urldef\tempurl%
\url{https://doi.org/10.18653/v1/2021.acl-long.416}
\showDOI{\tempurl}


\bibitem[Ng and Carley(2022)]%
        {ng2022botbuster}
\bibfield{author}{\bibinfo{person}{Lynnette Hui~Xian Ng} {and}
  \bibinfo{person}{Kathleen~M Carley}.} \bibinfo{year}{2022}\natexlab{}.
\newblock \showarticletitle{BotBuster: Multi-platform Bot Detection Using A
  Mixture of Experts}.
\newblock \bibinfo{journal}{\emph{arXiv preprint arXiv:2207.13658}}
  (\bibinfo{year}{2022}).
\newblock


\bibitem[Pastor-Galindo et~al\mbox{.}(2020)]%
        {pastor2020spotting}
\bibfield{author}{\bibinfo{person}{Javier Pastor-Galindo},
  \bibinfo{person}{Mattia Zago}, \bibinfo{person}{Pantaleone Nespoli},
  \bibinfo{person}{Sergio~L{\'o}pez Bernal}, \bibinfo{person}{Alberto~Huertas
  Celdr{\'a}n}, \bibinfo{person}{Manuel~Gil P{\'e}rez},
  \bibinfo{person}{Jos{\'e}~A Ruip{\'e}rez-Valiente},
  \bibinfo{person}{Gregorio~Mart{\'\i}nez P{\'e}rez}, {and}
  \bibinfo{person}{F{\'e}lix~G{\'o}mez M{\'a}rmol}.}
  \bibinfo{year}{2020}\natexlab{}.
\newblock \showarticletitle{Spotting political social bots in Twitter: A use
  case of the 2019 Spanish general election}.
\newblock \bibinfo{journal}{\emph{IEEE Transactions on Network and Service
  Management}} \bibinfo{volume}{17}, \bibinfo{number}{4}
  (\bibinfo{year}{2020}), \bibinfo{pages}{2156--2170}.
\newblock


\bibitem[Pedregosa et~al\mbox{.}(2011)]%
        {scikit-learn}
\bibfield{author}{\bibinfo{person}{F. Pedregosa}, \bibinfo{person}{G.
  Varoquaux}, \bibinfo{person}{A. Gramfort}, \bibinfo{person}{V. Michel},
  \bibinfo{person}{B. Thirion}, \bibinfo{person}{O. Grisel},
  \bibinfo{person}{M. Blondel}, \bibinfo{person}{P. Prettenhofer},
  \bibinfo{person}{R. Weiss}, \bibinfo{person}{V. Dubourg}, \bibinfo{person}{J.
  Vanderplas}, \bibinfo{person}{A. Passos}, \bibinfo{person}{D. Cournapeau},
  \bibinfo{person}{M. Brucher}, \bibinfo{person}{M. Perrot}, {and}
  \bibinfo{person}{E. Duchesnay}.} \bibinfo{year}{2011}\natexlab{}.
\newblock \showarticletitle{Scikit-learn: Machine Learning in {P}ython}.
\newblock \bibinfo{journal}{\emph{Journal of Machine Learning Research}}
  \bibinfo{volume}{12} (\bibinfo{year}{2011}), \bibinfo{pages}{2825--2830}.
\newblock


\bibitem[Peng et~al\mbox{.}(2020)]%
        {peng2020mixture}
\bibfield{author}{\bibinfo{person}{Hao Peng}, \bibinfo{person}{Roy Schwartz},
  \bibinfo{person}{Dianqi Li}, {and} \bibinfo{person}{Noah~A Smith}.}
  \bibinfo{year}{2020}\natexlab{}.
\newblock \showarticletitle{A Mixture of $ h-1$ Heads is Better than $ h $
  Heads}.
\newblock \bibinfo{journal}{\emph{arXiv preprint arXiv:2005.06537}}
  (\bibinfo{year}{2020}).
\newblock


\bibitem[Peng et~al\mbox{.}(2022)]%
        {peng2022domain}
\bibfield{author}{\bibinfo{person}{Huailiang Peng}, \bibinfo{person}{Yujun
  Zhang}, \bibinfo{person}{Hao Sun}, \bibinfo{person}{Xu Bai},
  \bibinfo{person}{Yangyang Li}, {and} \bibinfo{person}{Shuhai Wang}.}
  \bibinfo{year}{2022}\natexlab{}.
\newblock \showarticletitle{Domain-Aware Federated Social Bot Detection with
  Multi-Relational Graph Neural Networks}. In \bibinfo{booktitle}{\emph{2022
  International Joint Conference on Neural Networks (IJCNN)}}. IEEE,
  \bibinfo{pages}{1--8}.
\newblock


\bibitem[Pytorch(2018)]%
        {pytorch2018pytorch}
\bibfield{author}{\bibinfo{person}{Automatic Differentiation~In Pytorch}.}
  \bibinfo{year}{2018}\natexlab{}.
\newblock \bibinfo{title}{Pytorch}.
\newblock
\newblock


\bibitem[Raffel et~al\mbox{.}(2020)]%
        {raffel2020exploring}
\bibfield{author}{\bibinfo{person}{Colin Raffel}, \bibinfo{person}{Noam
  Shazeer}, \bibinfo{person}{Adam Roberts}, \bibinfo{person}{Katherine Lee},
  \bibinfo{person}{Sharan Narang}, \bibinfo{person}{Michael Matena},
  \bibinfo{person}{Yanqi Zhou}, \bibinfo{person}{Wei Li},
  \bibinfo{person}{Peter~J Liu}, {et~al\mbox{.}}}
  \bibinfo{year}{2020}\natexlab{}.
\newblock \showarticletitle{Exploring the limits of transfer learning with a
  unified text-to-text transformer.}
\newblock \bibinfo{journal}{\emph{J. Mach. Learn. Res.}} \bibinfo{volume}{21},
  \bibinfo{number}{140} (\bibinfo{year}{2020}), \bibinfo{pages}{1--67}.
\newblock


\bibitem[Rossi et~al\mbox{.}(2020)]%
        {rossi2020detecting}
\bibfield{author}{\bibinfo{person}{Sippo Rossi}, \bibinfo{person}{Matti Rossi},
  \bibinfo{person}{Bikesh Upreti}, {and} \bibinfo{person}{Yong Liu}.}
  \bibinfo{year}{2020}\natexlab{}.
\newblock \showarticletitle{Detecting political bots on Twitter during the 2019
  Finnish parliamentary election}. In \bibinfo{booktitle}{\emph{Proceedings of
  the 53rd Hawaii international conference on system sciences}}.
\newblock


\bibitem[Sayyadiharikandeh et~al\mbox{.}(2020)]%
        {sayyadiharikandeh2020detection}
\bibfield{author}{\bibinfo{person}{Mohsen Sayyadiharikandeh},
  \bibinfo{person}{Onur Varol}, \bibinfo{person}{Kai-Cheng Yang},
  \bibinfo{person}{Alessandro Flammini}, {and} \bibinfo{person}{Filippo
  Menczer}.} \bibinfo{year}{2020}\natexlab{}.
\newblock \showarticletitle{Detection of novel social bots by ensembles of
  specialized classifiers}. In \bibinfo{booktitle}{\emph{Proceedings of the
  29th ACM international conference on information \& knowledge management}}.
  \bibinfo{pages}{2725--2732}.
\newblock


\bibitem[Schlichtkrull et~al\mbox{.}(2018)]%
        {schlichtkrull2018modeling}
\bibfield{author}{\bibinfo{person}{Michael Schlichtkrull},
  \bibinfo{person}{Thomas~N Kipf}, \bibinfo{person}{Peter Bloem},
  \bibinfo{person}{Rianne van~den Berg}, \bibinfo{person}{Ivan Titov}, {and}
  \bibinfo{person}{Max Welling}.} \bibinfo{year}{2018}\natexlab{}.
\newblock \showarticletitle{Modeling relational data with graph convolutional
  networks}. In \bibinfo{booktitle}{\emph{European semantic web conference}}.
  Springer, \bibinfo{pages}{593--607}.
\newblock


\bibitem[Shazeer et~al\mbox{.}(2017)]%
        {shazeer2017outrageously}
\bibfield{author}{\bibinfo{person}{Noam Shazeer}, \bibinfo{person}{Azalia
  Mirhoseini}, \bibinfo{person}{Krzysztof Maziarz}, \bibinfo{person}{Andy
  Davis}, \bibinfo{person}{Quoc Le}, \bibinfo{person}{Geoffrey Hinton}, {and}
  \bibinfo{person}{Jeff Dean}.} \bibinfo{year}{2017}\natexlab{}.
\newblock \showarticletitle{Outrageously large neural networks: The
  sparsely-gated mixture-of-experts layer}.
\newblock \bibinfo{journal}{\emph{arXiv preprint arXiv:1701.06538}}
  (\bibinfo{year}{2017}).
\newblock


\bibitem[Shi et~al\mbox{.}(2023)]%
        {shi2023over}
\bibfield{author}{\bibinfo{person}{Shuhao Shi}, \bibinfo{person}{Kai Qiao},
  \bibinfo{person}{Jie Yang}, \bibinfo{person}{Baojie Song},
  \bibinfo{person}{Jian Chen}, {and} \bibinfo{person}{Bin Yan}.}
  \bibinfo{year}{2023}\natexlab{}.
\newblock \showarticletitle{Over-Sampling Strategy in Feature Space for Graphs
  based Class-imbalanced Bot Detection}.
\newblock \bibinfo{journal}{\emph{arXiv preprint arXiv:2302.06900}}
  (\bibinfo{year}{2023}).
\newblock


\bibitem[Silva et~al\mbox{.}(2019)]%
        {silva2019empirical}
\bibfield{author}{\bibinfo{person}{Andres~Garcia Silva},
  \bibinfo{person}{Cristian Berrio}, {and} \bibinfo{person}{Jos{\'e}~Manuel
  G{\'o}mez-P{\'e}rez}.} \bibinfo{year}{2019}\natexlab{}.
\newblock \showarticletitle{An empirical study on pre-trained embeddings and
  language models for bot detection}. In \bibinfo{booktitle}{\emph{Proceedings
  of the 4th Workshop on Representation Learning for NLP (RepL4NLP-2019)}}.
  \bibinfo{pages}{148--155}.
\newblock


\bibitem[Starbird(2019)]%
        {starbird2019disinformation}
\bibfield{author}{\bibinfo{person}{Kate Starbird}.}
  \bibinfo{year}{2019}\natexlab{}.
\newblock \showarticletitle{Disinformation's spread: bots, trolls and all of
  us}.
\newblock \bibinfo{journal}{\emph{Nature}} \bibinfo{volume}{571},
  \bibinfo{number}{7766} (\bibinfo{year}{2019}), \bibinfo{pages}{449--450}.
\newblock


\bibitem[Thieltges et~al\mbox{.}(2016)]%
        {thieltges2016devil}
\bibfield{author}{\bibinfo{person}{Andree Thieltges}, \bibinfo{person}{Florian
  Schmidt}, {and} \bibinfo{person}{Simon Hegelich}.}
  \bibinfo{year}{2016}\natexlab{}.
\newblock \showarticletitle{The devil’s triangle: Ethical considerations on
  developing bot detection methods}. In \bibinfo{booktitle}{\emph{2016 AAAI
  Spring Symposium Series}}.
\newblock


\bibitem[Uyheng and Carley(2020)]%
        {uyheng2020bots}
\bibfield{author}{\bibinfo{person}{Joshua Uyheng} {and}
  \bibinfo{person}{Kathleen~M Carley}.} \bibinfo{year}{2020}\natexlab{}.
\newblock \showarticletitle{Bots and online hate during the COVID-19 pandemic:
  case studies in the United States and the Philippines}.
\newblock \bibinfo{journal}{\emph{Journal of computational social science}}
  \bibinfo{volume}{3}, \bibinfo{number}{2} (\bibinfo{year}{2020}),
  \bibinfo{pages}{445--468}.
\newblock


\bibitem[Van~der Maaten and Hinton(2008)]%
        {van2008visualizing}
\bibfield{author}{\bibinfo{person}{Laurens Van~der Maaten} {and}
  \bibinfo{person}{Geoffrey Hinton}.} \bibinfo{year}{2008}\natexlab{}.
\newblock \showarticletitle{Visualizing data using t-SNE.}
\newblock \bibinfo{journal}{\emph{Journal of machine learning research}}
  \bibinfo{volume}{9}, \bibinfo{number}{11} (\bibinfo{year}{2008}).
\newblock


\bibitem[Varol et~al\mbox{.}(2017)]%
        {varol2017online}
\bibfield{author}{\bibinfo{person}{Onur Varol}, \bibinfo{person}{Emilio
  Ferrara}, \bibinfo{person}{Clayton Davis}, \bibinfo{person}{Filippo Menczer},
  {and} \bibinfo{person}{Alessandro Flammini}.}
  \bibinfo{year}{2017}\natexlab{}.
\newblock \showarticletitle{Online human-bot interactions: Detection,
  estimation, and characterization}. In \bibinfo{booktitle}{\emph{Proceedings
  of the international AAAI conference on web and social media}},
  Vol.~\bibinfo{volume}{11}. \bibinfo{pages}{280--289}.
\newblock


\bibitem[Wei and Nguyen(2019)]%
        {wei2019twitter}
\bibfield{author}{\bibinfo{person}{Feng Wei} {and} \bibinfo{person}{Uyen~Trang
  Nguyen}.} \bibinfo{year}{2019}\natexlab{}.
\newblock \showarticletitle{Twitter bot detection using bidirectional long
  short-term memory neural networks and word embeddings}. In
  \bibinfo{booktitle}{\emph{2019 First IEEE International Conference on Trust,
  Privacy and Security in Intelligent Systems and Applications (TPS-ISA)}}.
  IEEE, \bibinfo{pages}{101--109}.
\newblock


\bibitem[Wolf et~al\mbox{.}(2020)]%
        {wolf-etal-2020-transformers}
\bibfield{author}{\bibinfo{person}{Thomas Wolf}, \bibinfo{person}{Lysandre
  Debut}, \bibinfo{person}{Victor Sanh}, \bibinfo{person}{Julien Chaumond},
  \bibinfo{person}{Clement Delangue}, \bibinfo{person}{Anthony Moi},
  \bibinfo{person}{Pierric Cistac}, \bibinfo{person}{Tim Rault},
  \bibinfo{person}{Rémi Louf}, \bibinfo{person}{Morgan Funtowicz},
  \bibinfo{person}{Joe Davison}, \bibinfo{person}{Sam Shleifer},
  \bibinfo{person}{Patrick von Platen}, \bibinfo{person}{Clara Ma},
  \bibinfo{person}{Yacine Jernite}, \bibinfo{person}{Julien Plu},
  \bibinfo{person}{Canwen Xu}, \bibinfo{person}{Teven~Le Scao},
  \bibinfo{person}{Sylvain Gugger}, \bibinfo{person}{Mariama Drame},
  \bibinfo{person}{Quentin Lhoest}, {and} \bibinfo{person}{Alexander~M. Rush}.}
  \bibinfo{year}{2020}\natexlab{}.
\newblock \showarticletitle{Transformers: State-of-the-Art Natural Language
  Processing}. In \bibinfo{booktitle}{\emph{Proceedings of the 2020 Conference
  on Empirical Methods in Natural Language Processing: System Demonstrations}}.
  \bibinfo{publisher}{Association for Computational Linguistics},
  \bibinfo{address}{Online}, \bibinfo{pages}{38--45}.
\newblock
\urldef\tempurl%
\url{https://www.aclweb.org/anthology/2020.emnlp-demos.6}
\showURL{%
\tempurl}


\bibitem[Wu et~al\mbox{.}(2023a)]%
        {wu2023botshape}
\bibfield{author}{\bibinfo{person}{Jun Wu}, \bibinfo{person}{Xuesong Ye}, {and}
  \bibinfo{person}{Chengjie Mou}.} \bibinfo{year}{2023}\natexlab{a}.
\newblock \showarticletitle{BotShape: A Novel Social Bots Detection Approach
  via Behavioral Patterns}.
\newblock \bibinfo{journal}{\emph{arXiv preprint arXiv:2303.10214}}
  (\bibinfo{year}{2023}).
\newblock


\bibitem[Wu et~al\mbox{.}(2023b)]%
        {wu2023bottrinet}
\bibfield{author}{\bibinfo{person}{Jun Wu}, \bibinfo{person}{Xuesong Ye}, {and}
  \bibinfo{person}{Man~Yan Yuet}.} \bibinfo{year}{2023}\natexlab{b}.
\newblock \showarticletitle{BotTriNet: A Unified and Efficient Embedding for
  Social Bots Detection via Metric Learning}.
\newblock \bibinfo{journal}{\emph{arXiv preprint arXiv:2304.03144}}
  (\bibinfo{year}{2023}).
\newblock


\bibitem[Yan et~al\mbox{.}(2021)]%
        {doi:10.1177/1461444820942744}
\bibfield{author}{\bibinfo{person}{Harry~Yaojun Yan},
  \bibinfo{person}{Kai-Cheng Yang}, \bibinfo{person}{Filippo Menczer}, {and}
  \bibinfo{person}{James Shanahan}.} \bibinfo{year}{2021}\natexlab{}.
\newblock \showarticletitle{Asymmetrical perceptions of partisan political
  bots}.
\newblock \bibinfo{journal}{\emph{New Media \& Society}} \bibinfo{volume}{23},
  \bibinfo{number}{10} (\bibinfo{year}{2021}), \bibinfo{pages}{3016--3037}.
\newblock
\urldef\tempurl%
\url{https://doi.org/10.1177/1461444820942744}
\showDOI{\tempurl}
\showeprint{https://doi.org/10.1177/1461444820942744}


\bibitem[Yang et~al\mbox{.}(2022)]%
        {yang2022botometer}
\bibfield{author}{\bibinfo{person}{Kai-Cheng Yang}, \bibinfo{person}{Emilio
  Ferrara}, {and} \bibinfo{person}{Filippo Menczer}.}
  \bibinfo{year}{2022}\natexlab{}.
\newblock \showarticletitle{Botometer 101: Social bot practicum for
  computational social scientists}.
\newblock \bibinfo{journal}{\emph{arXiv preprint arXiv:2201.01608}}
  (\bibinfo{year}{2022}).
\newblock


\bibitem[Yang et~al\mbox{.}(2021)]%
        {yang2021covid}
\bibfield{author}{\bibinfo{person}{Kai-Cheng Yang}, \bibinfo{person}{Francesco
  Pierri}, \bibinfo{person}{Pik-Mai Hui}, \bibinfo{person}{David Axelrod},
  \bibinfo{person}{Christopher Torres-Lugo}, \bibinfo{person}{John Bryden},
  {and} \bibinfo{person}{Filippo Menczer}.} \bibinfo{year}{2021}\natexlab{}.
\newblock \showarticletitle{The covid-19 infodemic: Twitter versus facebook}.
\newblock \bibinfo{journal}{\emph{Big Data \& Society}} \bibinfo{volume}{8},
  \bibinfo{number}{1} (\bibinfo{year}{2021}),
  \bibinfo{pages}{20539517211013861}.
\newblock


\bibitem[Yang et~al\mbox{.}(2020)]%
        {yang2020scalable}
\bibfield{author}{\bibinfo{person}{Kai-Cheng Yang}, \bibinfo{person}{Onur
  Varol}, \bibinfo{person}{Pik-Mai Hui}, {and} \bibinfo{person}{Filippo
  Menczer}.} \bibinfo{year}{2020}\natexlab{}.
\newblock \showarticletitle{Scalable and generalizable social bot detection
  through data selection}. In \bibinfo{booktitle}{\emph{Proceedings of the AAAI
  conference on artificial intelligence}}, Vol.~\bibinfo{volume}{34}.
  \bibinfo{pages}{1096--1103}.
\newblock


\bibitem[Yang et~al\mbox{.}(2023)]%
        {yang2023fedack}
\bibfield{author}{\bibinfo{person}{Yingguang Yang}, \bibinfo{person}{Renyu
  Yang}, \bibinfo{person}{Hao Peng}, \bibinfo{person}{Yangyang Li},
  \bibinfo{person}{Tong Li}, \bibinfo{person}{Yong Liao}, {and}
  \bibinfo{person}{Pengyuan Zhou}.} \bibinfo{year}{2023}\natexlab{}.
\newblock \showarticletitle{FedACK: Federated Adversarial Contrastive Knowledge
  Distillation for Cross-Lingual and Cross-Model Social Bot Detection}.
\newblock \bibinfo{journal}{\emph{arXiv preprint arXiv:2303.07113}}
  (\bibinfo{year}{2023}).
\newblock


\bibitem[Zannettou et~al\mbox{.}(2019)]%
        {zannettou2019disinformation}
\bibfield{author}{\bibinfo{person}{Savvas Zannettou}, \bibinfo{person}{Tristan
  Caulfield}, \bibinfo{person}{Emiliano De~Cristofaro},
  \bibinfo{person}{Michael Sirivianos}, \bibinfo{person}{Gianluca Stringhini},
  {and} \bibinfo{person}{Jeremy Blackburn}.} \bibinfo{year}{2019}\natexlab{}.
\newblock \showarticletitle{Disinformation warfare: Understanding
  state-sponsored trolls on Twitter and their influence on the web}. In
  \bibinfo{booktitle}{\emph{Companion proceedings of the 2019 world wide web
  conference}}. \bibinfo{pages}{218--226}.
\newblock


\end{thebibliography}


\appendix





\end{document}